\documentclass[twocolumn,english,aps,prb,twocolum,superscriptaddress,bibnotes,amsmath,amssymb,floatfix]{revtex4-1}
\usepackage[colorlinks=true,citecolor=blue,linkcolor=magenta]{hyperref}

\usepackage[markup=nocolor, authormarkupposition=left]{changes} 
\usepackage{soul}
\usepackage[utf8]{inputenc}
\usepackage[english]{babel}
\usepackage{amsmath,amsfonts,amssymb}
\usepackage[T1]{fontenc}
\usepackage{url}

\usepackage{amsmath}
\usepackage{amsfonts}
\usepackage{amssymb}

\usepackage{epstopdf}
\usepackage{graphicx}
\usepackage{changes}

\usepackage{dcolumn}
\usepackage{tabularx}
\usepackage{booktabs}

\usepackage[font=small,labelfont=bf,justification=centerlast]{caption}

\begin{document}
%\title{Suppression of spatial mode interaction in ultralow-loss, \\multi-mode photonic integrated circuits}
\title{Compact, spatial-mode-interaction-free, ultralow-loss, \\nonlinear photonic integrated circuits}

\author{Xinru Ji}
\affiliation{Institute of Physics, Swiss Federal Institute of Technology Lausanne (EPFL), CH-1015 Lausanne, Switzerland}

\author{Junqiu Liu}
\email[]{junqiu.liu@epfl.ch}
\affiliation{Institute of Physics, Swiss Federal Institute of Technology Lausanne (EPFL), CH-1015 Lausanne, Switzerland}

\author{Jijun He}
\affiliation{Institute of Physics, Swiss Federal Institute of Technology Lausanne (EPFL), CH-1015 Lausanne, Switzerland}

\author{Rui Ning Wang}
\affiliation{Institute of Physics, Swiss Federal Institute of Technology Lausanne (EPFL), CH-1015 Lausanne, Switzerland}

\author{Zheru Qiu}
\affiliation{Institute of Physics, Swiss Federal Institute of Technology Lausanne (EPFL), CH-1015 Lausanne, Switzerland}

\author{Johann Riemensberger}
\affiliation{Institute of Physics, Swiss Federal Institute of Technology Lausanne (EPFL), CH-1015 Lausanne, Switzerland}

\author{Tobias J. Kippenberg}
\email[]{tobias.kippenberg@epfl.ch}
\affiliation{Institute of Physics, Swiss Federal Institute of Technology Lausanne (EPFL), CH-1015 Lausanne, Switzerland}

\maketitle

%%%%%%%%%%%%%%%%%%%%%%%%%%%%%%%%%%%%%%%%%%%%%%%%%%%%%%%%%%%%%%%%%%%%%%
%%%%%%%%%%%%%%%%%%%%%%%%%% A B S T R A C T %%%%%%%%%%%%%%%%%%%%%%%%%%%
%%%%%%%%%%%%%%%%%%%%%%%%%%%%%%%%%%%%%%%%%%%%%%%%%%%%%%%%%%%%%%%%%%%%%%

\noindent\textbf{Nonlinear photonics based on integrated circuits \cite{Moss:13, Eggleton:19} has enabled applications such as parametric amplifiers \cite{Foster:06}, soliton frequency combs \cite{Kippenberg:18}, supercontinua \cite{Gaeta:19}, and non-reciprocal devices \cite{Sounas:17}. 
Ultralow optical loss and the capability for dispersion engineering are essential, which necessitate the use of multi-mode waveguides.  
Despite that rich interaction among different spatial waveguide eigenmodes can give rise to novel nonlinear phenomena \cite{Yang:17a, Guo:17, Cole:17, Karpov:19, Liu:14, Xue:15, Matsko:16, Yang:16}, spatial mode interaction is typically undesired as it increases optical loss, perturbs local dispersion profile, and impedes soliton formation\cite{Herr:14a, KimC:21, Ramelow:14, Huang:16}. 
Adiabatic bends \cite{Chen:12, Cherchi:13}, such as Euler bends \cite{Fujisawa:17, Vogelbacher:19, Jiang:18} whose curvature varies linearly with their path length,  have been favoured to suppress spatial mode interaction. 
Adiabatic bends can essentially connect any two waveguide segments with adiabatic mode conversion \cite{Chen:12}, thus efficiently avoid mode mixing due to mode mismatch.
However, previous works lack quantitative measurement data and analysis to fairly evaluate the adiabaticity, and are not based on photonic integrated circuits with tight optical confinement and optical losses of few decibel per meter. 
Here, we adapt \cite{Fujisawa:17, Vogelbacher:19}, optimize, and implement Euler bends to build compact racetrack microresonators based on ultralow-loss, multi-mode, silicon nitride photonic integrated circuits \cite{Liu:21}. 
The racetrack microresonators feature a small footprint of only 0.21~mm$^2$ for 19.8 GHz free spectral range, critical for photonic integration of other functionalities, e.g. piezoelectric modulators \cite{Tian:20, Liu:20a} whose capacitance is proportional to the device area not perimeter.
We quantitatively investigate the suppression of spatial mode interaction in the racetrack microresonators with Euler bends, in comparison with normal racetracks using circular bends. 
We show that the optical loss rate (15.5~MHz) is preserved, on par with the mode interaction strength (25~MHz). 
This results in an unperturbed microresonator dispersion profile. 
We further demonstrate single dissipative Kerr soliton of 19.8~GHz repetition rate in the racetrack microresonator with Euler bends, without complex laser tuning schemes or auxiliary lasers. 
The optimized Euler bends and racetrack microresonators can be key building blocks for integrated nonlinear photonic systems \cite{Spencer:18, Newman:19, Liu:20, Zhang:19}, as well as linear circuits for programmable processors \cite{Zhuang:15} and photonic quantum computing \cite{WangJW:20, Arrazola:21}. 
}

%%%%%%%%%%%%%%%%%%%%%%%
%%%%% Introduction %%%%%
%%%%%%%%%%%%%%%%%%%%%%%

Over the past decade, integrated photonic technology has been successfully translated from laboratory research into industrial applications \cite{Rickman:14}, and has revolutionized today's optical communication networks \cite{Thomson:16, Agrell:16}. 
Besides silicon -- the backbone microelectronic material,  many new material platforms have been developed, as well as heterogeneous and hybrid integration techniques \cite{Komljenovic:16, Kaur:21, Margalit:21} to merge them with silicon. 
Among these materials, amorphous silicon nitride (Si$_3$N$_4$) \cite{Moss:13, Munoz:19} -- first proposed in 1980's \cite{Henry:87} for integrated photonics due to its wide transparency window from visible to mid-infrared and a wide bandgap of 5 eV -- has made significant progress recently. 
Stoichiometric Si$_3$N$_4$ can be deposited via chemical vapor deposition (CVD) and is CMOS-compatible. 
Recent advances in fabrication \cite{Xuan:16, Liu:21, Ji:21,Ye:19b} have achieved Si$_3$N$_4$ photonic integrated circuits (PIC) that are free from cracks and feature tight optical confinement, high power handling capability, wideband engineering of anomalous group velocity dispersion (GVD) \cite{Okawachi:14}, and ultralow optical losses near 1 dB/m \cite{Liu:21, Ye:21}. % with a strong third order nonlinear coefficient of 1 W$^{-1}$m$^{-1}$. 
Combining with its high Kerr nonlinearity, weak Brillouin scattering \cite{Gyger:20} and negligible two-photon absorption, Si$_3$N$_4$ has become the material of choice particularly for Kerr nonlinear photonics, such as microresonator soliton frequency combs (``soliton microcombs'') \cite{Kippenberg:18}, chip-based supercontinua\cite{Gaeta:19}, and ultralow-threshold optical parametric oscillators \cite{Li:16, Lu:19}. 

\begin{figure*}[t!]
\includegraphics[width=0.95\linewidth]{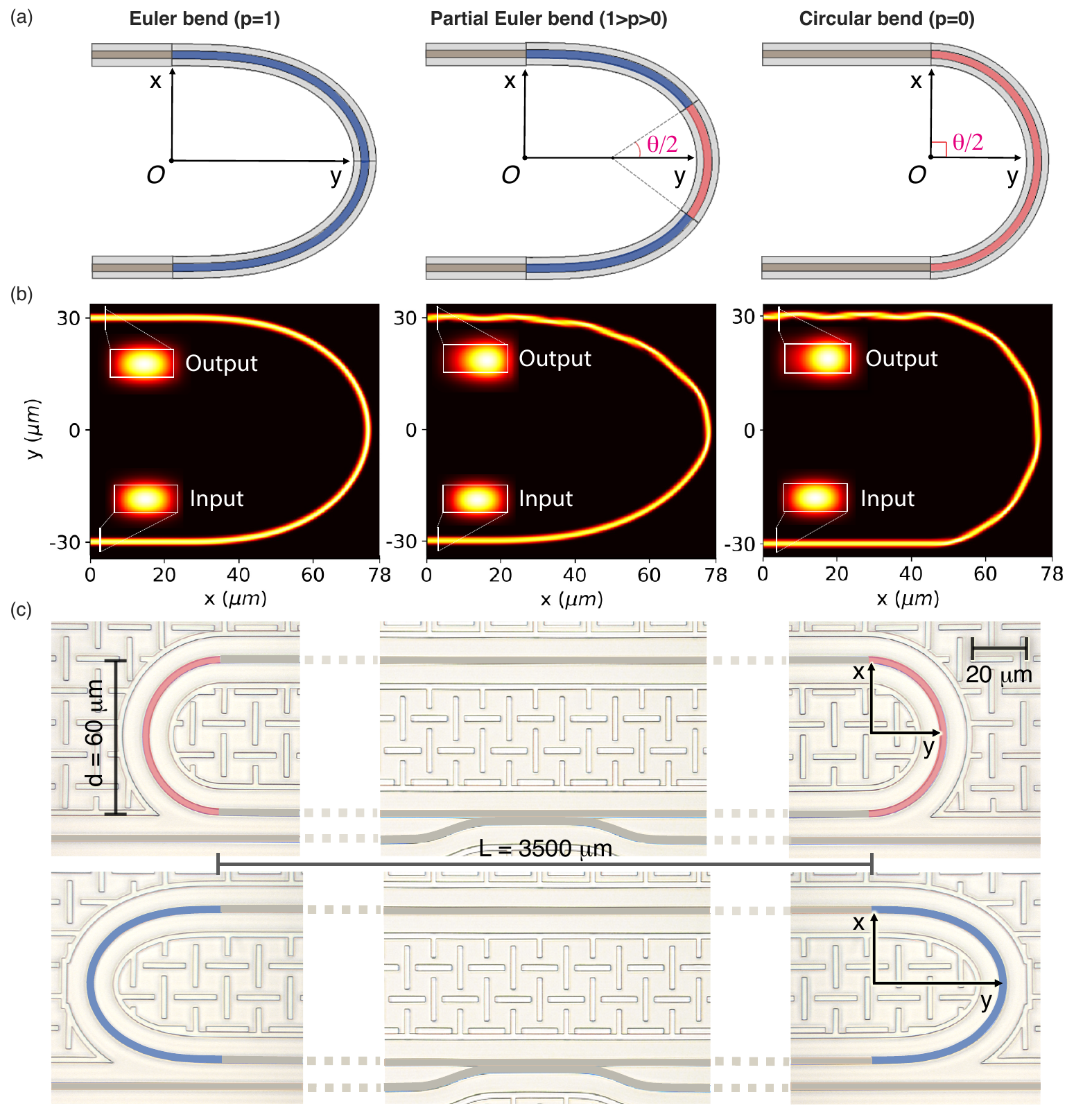}
\caption{
\textbf{Schematics, designs and simulations of Euler bends for integrated Si$_3$N$_4$ racetrack microresonators.}
(a) Illustration of the difference between Euler bends (blue) and circular bends (red) in Cartesian coordinate system $x$ -- $y$, to connect two straight waveguides (brown) in the racetrack microresonator within a $\pi$-bend (180$^\circ$~bend). 
The Euler portion factor $p$, which depends on the circular angle $\theta$, denotes the ratio of Euler bends (blue) in the total $\pi$-bend. 
Here, $p=1$ ($\theta=0$), $1>p>0$ ($\pi>\theta>0$), and $p=0$ ($\theta=\pi$) corresponds to Euler bend,  partial Euler bend, and circular Euler bend, respectively.  
(b) FDTD simulations of mode propagation in three different bends, showing that higher $p$ offers better adiabatic mode conversion. 
In each case, the TE$_{00}$ mode of the straight waveguide is launched and propagates through the $\pi$-bend. 
Mode mixing during propagation is revealed by comparing the input and output modes, as well as the overall mode propagation profiles. 
It is observed that the launched TE$_{00}$ mode is preserved in the Euler bend ($p=1$), while it experiences distortion in the partial Euler bend ($p=0.6$) and circular bend ($p=0$). 
(c) Optical microscope images showing fabricated Si$_3$N$_4$ racetrack microresonators with Euler bends ($p=1$) and circular bends ($p=0$). 
Both racetrack microresonators have the same span of $d=60$ $\mu$m for the $\pi$-bends, and length of $L=3500$ $\mu$m. 
The device footprint is approximately 0.21 mm$^2$. 
}
\label{Fig:1}
\vspace{-0.4cm}
\end{figure*}

Multi-mode waveguides with tight optical confinement are widely used in integrated nonlinear photonics. 
The large waveguide cross-sections reduce optical loss by reducing light interaction with waveguide surface roughness \cite{Zhang:20}.
Indeed, state-of-the-art integrated waveguides of losses near 1~dB/m are all multi-mode \cite{Xuan:16, Ji:21, Liu:21, Ye:21, Zhang:17}.
In addition, stimulated Brillouin scattering \cite{Eggleton:19, Poulton:12} and indirect-interband transition \cite{Sounas:17, Yu:09} among different spatial modes are the basis of magnetic-free non-reciprocal devices \cite{Lira:12, Sohn:18, Kittlaus:18, Tian:21a}.
Equally important, harmonic generation is facilitated by employing phase matching among different spatial modes \cite{Levy:11, Surya:18, Porcel:17, Lu:21, Nitiss:21}.
For Kerr nonlinear photonics, anomalous GVD is required, which is realized via geometric dispersion engineering \cite{Okawachi:14} using multi-mode waveguides.
Novel phenomena can arise from spatial mode interaction in multi-mode microresonators, such as the generation of Stokes solitons \cite{Yang:17a}, breathing solitons \cite{Guo:17}, soliton crystals \cite{Cole:17, Karpov:19}, dark pulses \cite{Liu:14, Xue:15}, and spatial-mode-interaction-induced dispersive waves \cite{Matsko:16, Yang:16}. 
Single avoided-mode crossing (AMX) resulting from strong mode coupling can facilitate the stabilization of bright solitons \cite{Bao:17, Yi:17} or the generation of dark pulses \cite{Xue:15a, KimB:19}.
However, spatial mode coupling in highly over-moded waveguides is uncontrollable, yielding spectrally distributed AMXs with random strengths. 
These coexistent AMXs can often prohibit soliton formation \cite{Herr:14a, KimC:21} and significantly distort soliton spectra \cite{Ramelow:14, Huang:16}.
Therefore, suppression of spatial mode coupling in multi-mode waveguides is still desired for stable and reliable access to the single soliton regime \cite{Liu:20}.

In multi-mode waveguides, defects and surface roughness can introduce spatial mode interaction.
In addition, mode mixing occurs when waveguide curvature experiences non-adiabatic transition, e.g. from a straight section to a curved section, due to bending-induced mode mismatch.
Despite that the waveguide cross-section is identical, the waveguide bending causes changes of waveguide eigenmodes due to the modified effective indices \cite{Subramaniam:97}, and finally yields spatial mode mixing. % if the mode conversion is non-adiabatic. 
In ultralow-loss PICs where defects and surface roughness are carefully addressed, mode mixing due to non-adiabatic transition can be the dominant cause for spatial mode interaction. 
Such mode interaction represents a major loss channel \cite{Pfeiffer:17b}, and can severely limit the performance of ultralow-loss PICs and planar microresonators.

%%%%%%%%%%%%%%%%%%%%%%%
%%%%% Euler bend %%%%%
%%%%%%%%%%%%%%%%%%%%%%%

In this work, we demonstrate a novel photonic design that suppresses spatial mode interaction and effectively operates multi-mode waveguides in the single-mode regime. 
Using a racetrack microresonator that features high quality factor $Q>10^7$ and consists of two straight waveguides and two adiabatic Euler bends, we qualitatively characterize the suppressed spatial mode interaction. 

Racetrack microresonators of tens of gigahertz free spectral range (FSR) are widely used for the generation of soliton microcombs or electro-optic combs with microwave repetition rates \cite{Liu:20, Zhang:19}.  
A normal racetrack microresonator consists of two straight waveguides and two circular bends.
Since the optical eigenmodes are modified due to the waveguide curvature, the eigenmode bases are different for the straight waveguide (with an infinite bending radius) and curved waveguide (with a finite bending radius). 
Therefore, mode mixing is introduced due to the abrupt mode transition. 
In high-$Q$ racetrack microresonators where light circulates many round-trips, such mode mixing can severely impact device performance, increase losses, and shift resonance eigen-frequencies. 
For example, several studies have highlighted difficulties to access soliton states of microwave repetition rates in racetrack or finger-shape microresonators \cite{Johnson:12, Xuan:16, Kim:17}. 
To avoid mode mixing, microring resonators are used and single solitons with repetition rates in the microwave X- and K-band are generated \cite{Liu:20}, however the devices suffer from significant footprints (larger than 4 mm$^2$) and constrained device density. 
Alternatively, mode filtering elements \cite{Kordts:16, KimC:21} can be used to suppress mode mixing. 

%%%%%%%%%%%%%%%%%%%%%%%
%%%%% Design %%%%%
%%%%%%%%%%%%%%%%%%%%%%%

Here, we optimize racetrack microresonators by replacing the two circular bends with two Euler bends \cite{Fujisawa:17, Vogelbacher:19} that have adiabatic radius transition from infinite (in the straight section) to finite (in the bending section). 
An Euler bend has a curvature (the inverse of radius) varying linearly with its path length. 
Adiabatic bends, including Euler bends, have profound applications in railroad and highway engineering, and have later been used in integrated photonics.
For example, a general design for adiabatic waveguide connection \cite{Chen:12} has been applied to create ultralow-loss, meter-long, suspended silica photonic delay lines \cite{Lee:12a}. 
Modified Euler bends \cite{Jiang:18} have been applied on multi-mode silicon-on-insulator (SOI) racetrack microresonators \cite{Zhang:20} to achieve high $Q$. 
Nonlinear adiabatic bends \cite{Ji:21} have been used on Si$_3$N$_4$ microring resonators to to achieve low loss and broadband external coupling with bus waveguides. 
 
Figure \ref{Fig:1} highlights the difference between circular bends and adiabatic Euler bends. 
In Cartesian coordinate system $x$ -- $y$, the circular bend can be expressed as $x^2+y^2=R^2$, where $R$ is the constant bending radius. 
In comparison, the Euler bend, illustrated in Ref. \cite{Vogelbacher:19}, can be expressed as 
\begin{equation}
\begin{aligned}
&x(s)=\int_{0}^{s} \cos \left(\frac{\alpha}{2} \cdot u^{2}\right) \text{d}u \\
&y(s)=\int_{0}^{s} \sin \left(\frac{\alpha}{2} \cdot u^{2}\right) \text{d}u
\end{aligned}
\end{equation}
where $\alpha$ is the linear rate of curvature ($k$) change with path length ($s$), i.e. $k(s)=\alpha\cdot s$. 
The curvature $k$ determines the local bending radius, i.e. $R(s)=[\text{d}\theta(s)/\text{d}s]^{-1}=k(s)^{-1}$.  

To connect two straight waveguides separated by distance $d$ in the racetrack microresonator,  a $\pi$-bend (i.e. 180$^\circ$~bend) consisting of two Euler bend sections and a circular bend section can be used, as illustrated in Fig. \ref{Fig:1}(a) middle. 
The two Euler bend sections connect each straight waveguide to the circular bend of angle $\theta\in(0, \pi)$, allowing for adiabatic mode conversion from infinite radius ($R=\infty$) to a finite value $R_p=(\alpha\pi p)^{-1/2}$, and vice versa. 
Here, the Euler portion factor $p$,  denoting the ratio of two Euler bends in the total $\pi$-bend, is defined as 
\begin{equation}
p=1-\frac{\theta}{\pi}
\end{equation}
We emphasize that, when the values of $p$ and $d$ are given, the linear curvature changing rate $\alpha$ is uniquely determined,  as
\begin{equation}
\alpha=\frac{[2\sqrt{2}\int_{0}^{\sqrt{\pi p/2}} \sin t^2 \text{d}t+\frac{2}{\sqrt{\pi p}}\sin \frac{\pi(1-p)}{2}]^2}{d^2}
\label{Eq:alpha}
\end{equation}
Supplementary Information shows the derivation and numerical plot of $\alpha$ as a function of $d$ and $p$. 
Particularly, the numerical plot of $\alpha$ as a function of $p$ with a constant $d$ shows that, the value of $\alpha$ decreases monotonously with $p$, and its minimum value is reached when $p=1$. 
This may suggest that $p=1$ gives highest adiabaticity, as the linear rate $\alpha$ of bending curvature change is smallest. 

%%%%%%%%%%%%%%%%%%%%%%%%%
\begin{figure*}[t!]
\includegraphics[width=\linewidth]{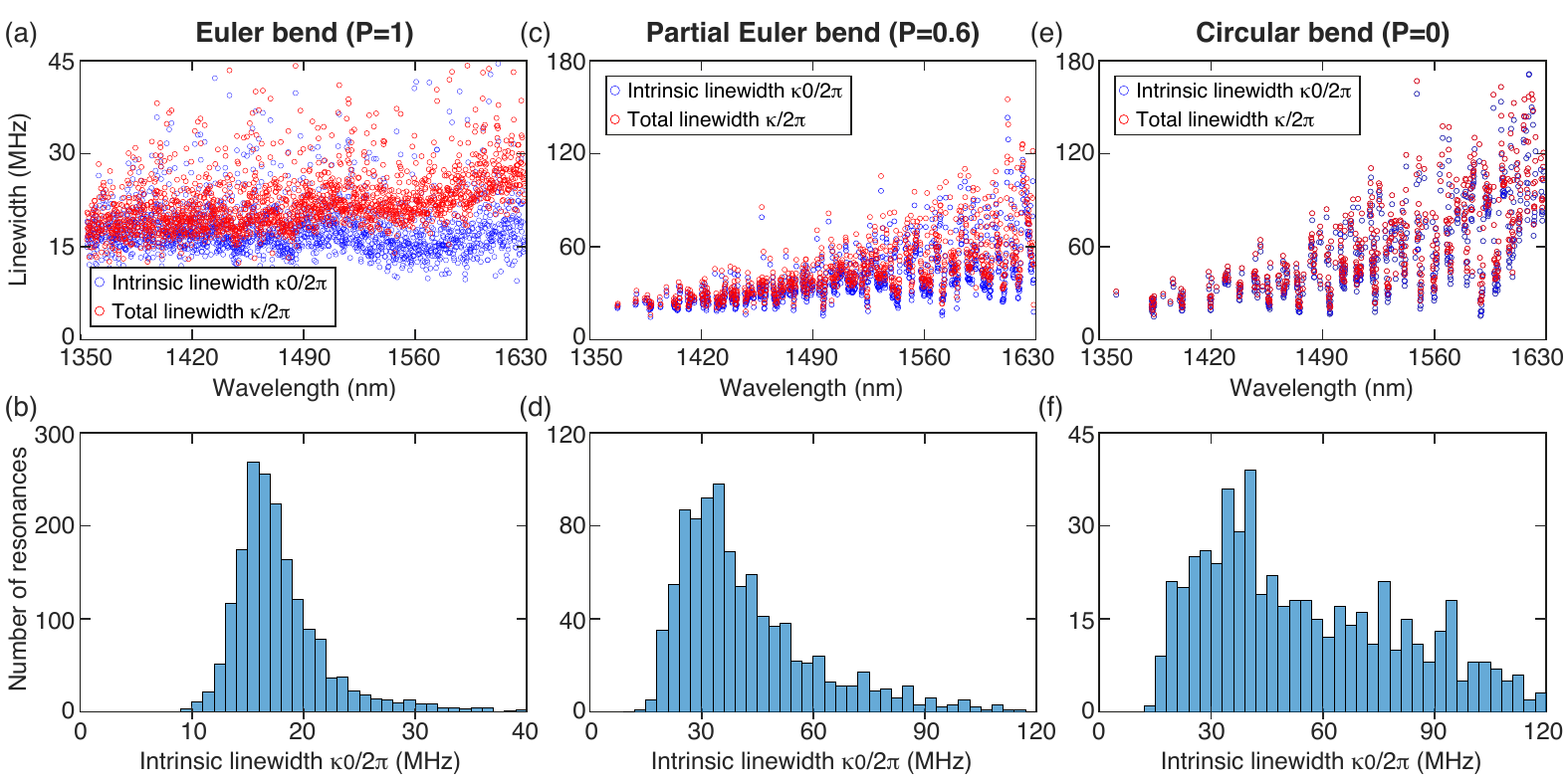}
\caption{
\textbf{Resonance linewidth characterization of racetrack microresonators with different bends}. 
Measured intrinsic linewidth $\kappa_0/2\pi$ and loaded linewidth $\kappa/2\pi$, and histogram of $\kappa_0/2\pi$ for the racetrack microresonators with Euler bends (a,b), partial Euler bends (c, d), and circular bends (e, f). 
}
\label{Fig:2}
\vspace{-0.3cm}
\end{figure*}
%%%%%%%%%%%%%%%%%%%%%%%%%

Figure~\ref{Fig:1}(a) illustrates the difference between a full Euler bend ($\theta=0, p=1$), a partial Euler bend ($p\in(0, 1)$), and a circular bend ($\theta=\pi, p=0$).  
To illustrate adiabatic mode conversion, finite-difference time-domain (FDTD) numerical simulations of mode propagation \cite{Pfeiffer:17b} in the three different waveguide bends are performed. 
In each case, the fundamental transverse-electric (TE$_{00}$) mode of the straight waveguide is launched and propagates through the bend. 
The mode propagation profile, as well as the output waveguide mode in the other straight waveguide, is recorded. 
In our simulation model, the waveguide cross-sections is 2.2~$\mu$m width and 0.95~$\mu$m height, and the two straight waveguides are separated by $d=60$~$\mu$m to each other.  
This waveguide cross-section supports ten spatial eigenmodes in total (see Supplementary Information). 
As shown in Fig.~\ref{Fig:1}(b), the launched TE$_{00}$ mode is preserved in the Euler bend ($p=1$), while it experiences considerable coupling to other waveguide modes in the partial Euler bend ($p=0.6$) and circular bend ($p=0$). 
In the latter cases, the fundamental TE$_{00}$ mode is predominantly coupled to the first-order transverse-electric (TE$_{10}$) mode.  
These two modes interfere during co-propagation, causing characteristic oscillations in the mode propagation profile in the waveguide. 

%However, we note that, all previous works lacks sufficient resonance statistics to quantitatively evaluate the mode coupling strength over the entire telecommunication band.  In addition, the optical losses in these works are insufficient low to observed spatial-mode-coupling induced loss at dB/m level. 
%\cite{Cherchi:13, Fujisawa:17}

%%%%%%%%%%%%%%%%%%%%%%%
%%%%% Q Measurement %%%%%
%%%%%%%%%%%%%%%%%%%%%%%

Next, we experimentally study and compare racetrack microresonators with three different bends.  
We fabricated Si$_3$N$_4$ racetrack microresonators using the photonic Damascene reflow process \cite{Liu:21}. 
Three different types of racetrack microresonators are fabricated, which have Euler bends ($p=1$), partial Euler bends ($p=0.6$) and circular bends ($p=0$), as shown in Fig. \ref{Fig:1}.  
We emphasize that, these three racetrack microresonators are fabricated on the same photonic chip of $5\times5$ mm$^2$, and are separated by 400 $\mu$m in a row (see design layout in Supplementary Information). 
This is to minimize fabrication impact on device performance due to the parameter variation over the 4-inch wafer scale. 
The Si$_3$N$_4$ waveguide has a cross-section of 2.2~$\mu$m width and 0.90 $\mu$m height. 
All racetrack microresonators have the same span of $d=60$~$\mu$m for the $\pi$-bends, and length of $L=3500$~$\mu$m for the straight waveguides, to minimize the device footprint (0.21~mm$^2$). 
The 60~$\mu$m span used in our Si$_3$N$_4$ platform is same as that used in high-$Q$ SOI racetrack microresonators \cite{Zhang:20}, despite that SOI waveguides have higher mode confinement. 
Figure~\ref{Fig:1}(c) shows the microscope images and aspect ratios of racetrack microresonators with Euler bends ($p=1$) or circular bends ($p=0$). 

%%%%%%%%%%%%%%%%%%%%%%%%%
\begin{figure*}[t!]
\includegraphics[width=\linewidth]{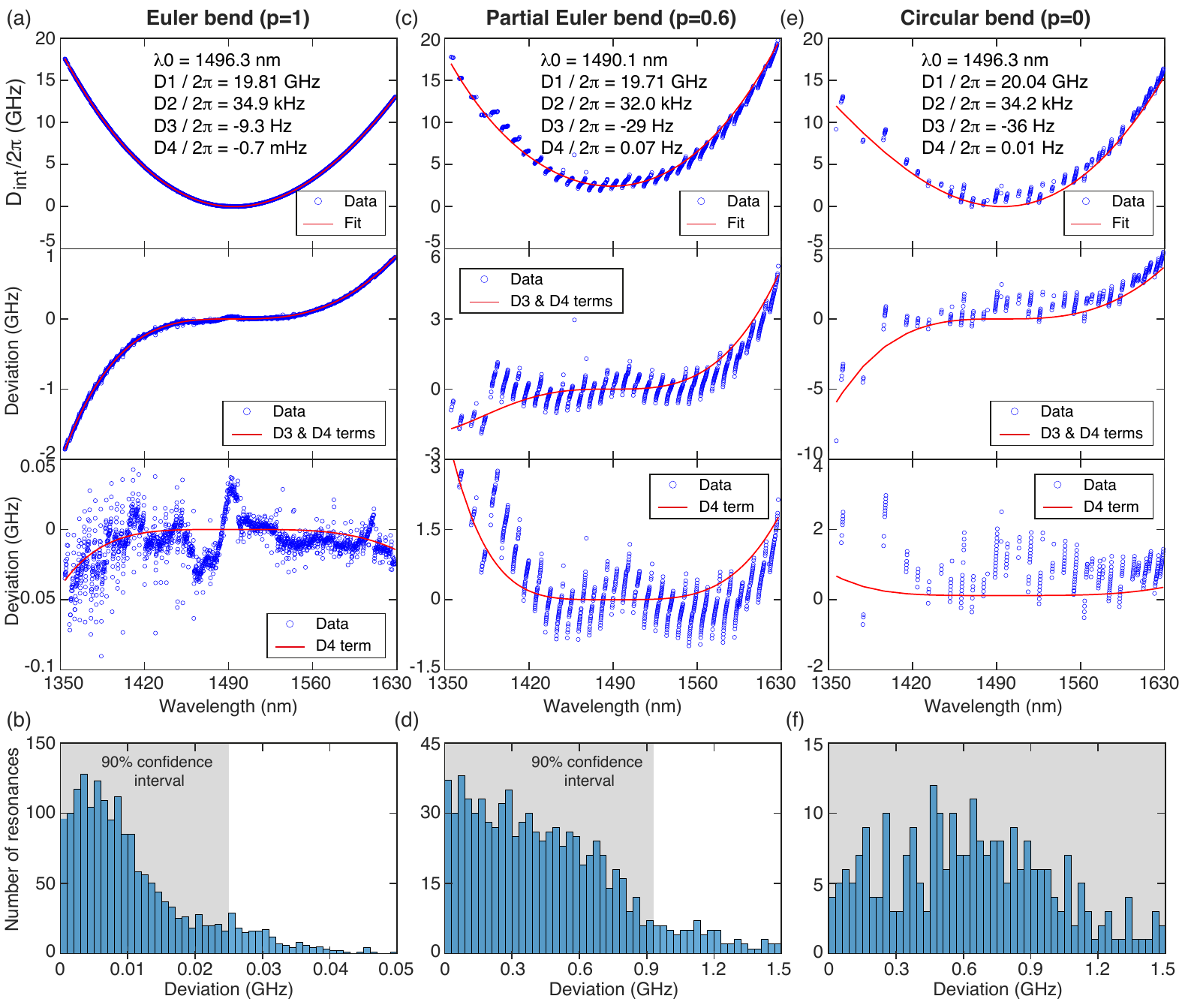}
\caption{
\textbf{Integrated dispersion characterization of racetrack microresonators with different bends.}
The measured integrated microresonator dispersion fitted with $D_\text{int}(\mu)=D_2\mu^2/2+D_3\mu^3/6+D_4\mu^4/24$, the resonance frequency deviations from $D_3\mu^3/6+D_4\mu^4/24$ and $D_4\mu^4/24$, and the histogram of frequency deviations from the $D_4\mu^4/24$ curve, for the racetrack microresonators with Euler bends (a,b), partial Euler bends (c, d), and circular bends (e, f). 
Values of each dispersion terms are marked in (a, c, e). 
Grey-shaded areas in (b, d, f) mark the 90\% confidence intervals. 
}
\label{Fig:3}
\vspace{-0.3cm}
\end{figure*}
%%%%%%%%%%%%%%%%%%%%%%%%%

To characterize the resonance linewidths (i. e. microresonator loss) and microresonator dispersion, we use frequency-comb-assisted, cascaded diode laser spectroscopy, to cover the entire telecommunication E- to L-band (1350 to 1630~nm) \cite{Liu:16}.  
When light is coupled into the microresonator and the laser frequency scans continuously, the optical transmission spectrum of the microresonator is acquired (see Supplementary Information).  
The instantaneous laser frequency of each recorded data point is calibrated by beating the laser with a commercial, self-referenced, fiber-laser-based optical frequency comb \cite{DelHaye:09}. %, further assisted with a Mach-Zehnder interferometer\cite{Pfeiffer:18b}. 
The microresonator transmission spectrum is further referenced to a molecular absorption spectroscopy to extract the absolute laser frequency offset. 
This method allows to identify each resonance and obtain the precise resonance frequency $\omega/2\pi$. 
By fitting each resonance profile \cite{Li:13}, the intrinsic loss $\kappa_0/2\pi$, bus-waveguide-to-microresonator external coupling strength $\kappa_\text{ex}/2\pi$, and loaded (total) linewidth $\kappa/2\pi=(\kappa_0+\kappa_\text{ex})/2\pi$ are extracted for each resonance. 

Here we mainly focus on the TE$_{00}$ mode, and investigate mode interaction of the TE$_{00}$ mode with other spatial modes. 
First, we study the intrinsic loss $\kappa_0/2\pi$ and loaded linewidth $\kappa/2\pi$ of each resonance from the three different racetrack microresonators. 
Figure~\ref{Fig:2}(a, c, e) reveal the wavelength dependence of $\kappa_0/2\pi$ and $\kappa/2\pi$. 
Figure~\ref{Fig:2}(b, d, f) plot the histogram of measured $\kappa_0/2\pi$ values from Fig.~\ref{Fig:2}(a, c, e), respectively. 
For the racetrack microresonator with Euler bends ($p=1$) shown in Fig. \ref{Fig:2}(a, b), no prominent wavelength dependence of $\kappa_0/2\pi$ is observed, and the most probable value in the histogram is $\kappa_0/2\pi=15.5$ MHz, corresponding to $2.4$ dB/m linear loss and a microresonator intrinsic quality factor $Q_0\approx13\times10^6$. 
However, for the racetrack microresonator with partial Euler bends ($p=0.6$) shown in Fig.~\ref{Fig:2}(c), higher $\kappa_0/2\pi$ with longer wavelength is observed, as well as spectrally periodic, vertical striations caused by inter-modal interference (see Supplementary Information). 
The most probable value of $\kappa_0/2\pi$ is increased to $34.5$ MHz as shown in Fig.~\ref{Fig:2}(d). 
Both observations are due to the interaction of the TE$_{00}$ mode with other spatial modes. 
Since the eigenmode mismatch between the straight waveguide and the circular bend is larger for longer wavelength, the spatial mode interaction is stronger and causes higher loss.  
These two observations are further verified in the racetrack microresonator with circular bends ($p=0$) shown in Fig.~\ref{Fig:2}(e, f). 
Here, spatial mode interaction is so strong that many resonances experience greatly increased $\kappa_0/2\pi$,  much higher than $\kappa_\text{ex}/2\pi$. 
Therefore, these resonances are strongly under-coupled \cite{Cai:00} and cannot be resolved in the transmission spectrum (see Supplementary Information).  
 
%%%%%%%%%%%%%%%%%%%%%%%
%%%%% Dispersion Measurement %%%%%
%%%%%%%%%%%%%%%%%%%%%%%

To quantitatively characterize the strength of spatial mode interaction, we further study the microresonator dispersion profile, and investigate AMXs. 
Figure \ref{Fig:3}(a, c, e) top panels show the measured integrated microresonator dispersion fitted with 
\begin{equation}
D_\text{int}(\mu)=\omega_{\mu}-\omega_0-D_1\mu=D_2\mu^2/2+D_3\mu^3/6+D_4\mu^4/24
\end{equation}
where $\omega_{\mu}/2\pi$ is the $\mu$-th resonance frequency relative to the reference resonance frequency $\omega_0/2\pi$ (wavelength $\lambda_0$), $D_1/2\pi$ corresponds to microresonator FSR, $D_2/2\pi$ is GVD, and $D_3$ and $D_4$ are higher-order dispersion terms. 
To reveal AMXs, $D_2$ and $D_3$ are removed from $D_\text{int}$. % as shown in Fig. \ref{Fig:3}(a, c, e) middle and bottom panels, respectively. 
Figure \ref{Fig:3}(a, c, e) middle and bottom panels show, respectively, 
\begin{equation}
\begin{aligned}
D_\text{int}-D_2\mu^2/2&=D_3\mu^3/6+D_4\mu^4/24 \\
D_\text{int}-D_2\mu^2/2-D_3\mu^3/6&=D_4\mu^4/24
\end{aligned}
\end{equation} 
%Figure \ref{Fig:3}(a, c, e) middle panels show $D_\text{int}-D_2\mu^2/2=D_3\mu^3/6+D_4\mu^4/24$. 
%Figure \ref{Fig:3}(a, c, e) bottom panels show $D_\text{int}-D_2\mu^2/2-D_3\mu^3/6=D_4\mu^4/24$. 

For the racetrack microresonator with Euler bends ($p=1$), AMXs are only revealed when $D_2$ and $D_3$ terms are both removed. 
The histogram of resonance frequency deviations from the $D_4\mu^4/24$ curve is plotted in Fig.~\ref{Fig:3}(b), with 90\% confidence interval below 25~MHz, i.e. 90\% of the total analyzed resonances have frequency deviations below 25~MHz. 
This value is on par with the loaded resonance linewidth (i.e. total photon loss rate) shown in Fig.~\ref{Fig:2}(a). 
%The area of this racetrack is 0.21 mm$^2$. 
The residual AMXs revealed in Fig.~\ref{Fig:3}(a) bottom might also originate from the bus waveguide coupling section which has been revealed in FDTD simulations in Ref. \cite{Pfeiffer:17b}. 
The bus waveguide coupling section can be optimized using asymmetric directional couplers \cite{Chin:98}.

In comparison, for the racetrack microresonator with partial Euler bends ($p=0.6$), AMXs are already revealed by $D_\text{int}$ fit in Fig.~\ref{Fig:3}(c) top. 
The enhanced AMXs lead to inaccurate fit and extraction of $D_3$ and $D_4$ values.  
Still, following the same procedure, Fig.~\ref{Fig:3}(d) shows the histogram of resonance frequency deviations from the $D_4\mu^4/24$ curve, with 90\% confidence interval below 930 MHz.  
Another two pairs of racetrack microresonators with $p=1$ and $p=0.6$ (four devices in total) are shown in Supplementary Information,  and present similar trends in microresonator dispersion profiles.  

Furthermore, for the racetrack microresonator with circular bends ($p=0$), the missing resonances and exaggerated AMXs prohibit to fit $D_\text{int}$ in Fig.~\ref{Fig:3}(e) top, leading to infeasibility to extract $D_3$ and $D_4$ values. 
Many resonances have frequency deviations more than 1~GHz, as shown in Fig. \ref{Fig:3}(f). 
We emphasize that, such exaggerated AMXs are caused by the circular bends of only 60~$\mu$m diameter for 19.8~GHz FSR. 
Previous work \cite{Anderson:21} has shown single soliton generation in racetrack microresonators driven by picosecond optical pulses. 
However, there, the diameter of the circular bend is 400~$\mu$m for 28~GHz FSR and 0.6~mm$^2$ device footprint (see design layout in Supplementary Information), thus the eigenmode mismatch between the straight waveguide and the circular bend is much smaller than that in our current case, leading to weaker mode coupling. 

%%%%%%%%%%%%%%%%%%%%%%%
%%%%%%% Soliton %%%%%%%
%%%%%%%%%%%%%%%%%%%%%%%

\begin{figure*}[t!]
\includegraphics[width=\linewidth]{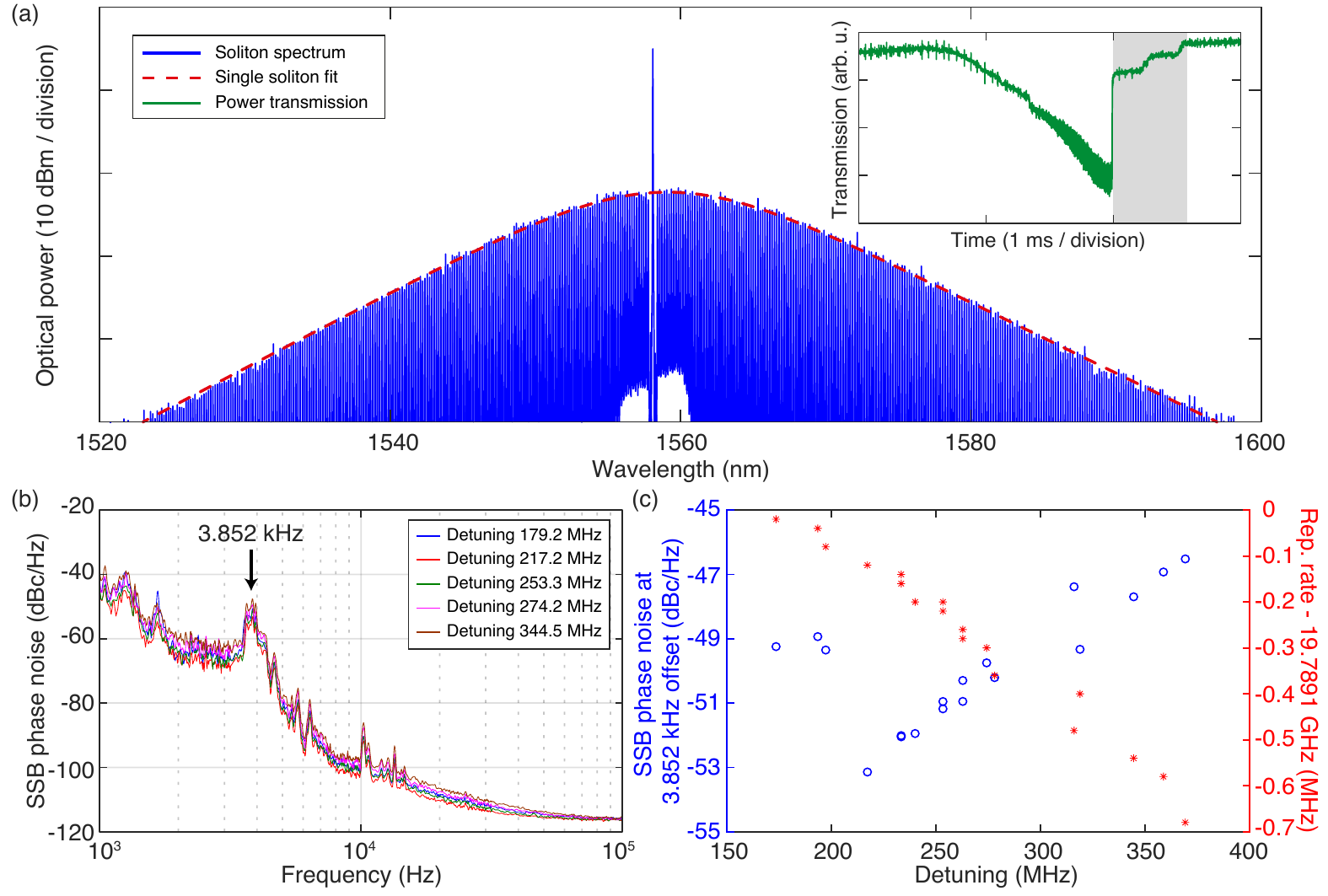}
\caption{
\textbf{Single soliton generation in the racetrack microresonator with Euler bends, and the phase noise characterization of soliton repetition rate}.  
(a) Single soliton spectra of 19.8~GHz repetition rate. 
No prominent dispersive wave features caused by AMXs are observed. 
The single soliton spectrum fit shows a 3-dB bandwidth of 16.3~nm, corresponding to a pulse duration of 156~fs. 
The estimated on-chip CW pump power is 55.7~mW.
Note that a band-pass filter is used to filter out the EDFA’s amplified spontaneous emission noise in the pump laser, and a fiber Bragg grating is used to filter out the pump laser in the soliton spectra. 
Inset: When the pump laser scans across the resonance, a soliton step of $\sim0.5$ ms length in the microresonator transmission is seen (marked in the gray zoom).
(b) SSB phase noise measurement with different soliton detuning values. 
No prominent phase noise change due to the quiet point operation is observed. 
The feature at 3.852 kHz Fourier offset frequency is caused by the diode laser pump \cite{Liu:20}. 
(c) SSB phase noise at 3.852 kHz Fourier offset frequency and measured repetition rate shift with different soliton detuning values. 
The absence of strong quiet point is likely due to the inhibited dispersive wave generation that is caused by the suppressed spatial mode interaction and avoided-mode crossings. 
}
\label{Fig:4}
\vspace{-0.3cm}
\end{figure*}

Finally, we demonstrate single soliton generation of 19.8 GHz repetition rate in the racetrack microresonator with Euler bends ($p=1$). 
When the continuous-wave (CW) pump laser scans across the resonance from the blue-detuned side to the red-detuned side, a step indicating soliton formation \cite{Herr:14} in the microresonator transmission spectrum is observed. 
Here we observe a soliton step length of $\sim0.5$ ms, on par with the previously reported value in a 100-GHz-FSR microresonator \cite{Liu:18a}. 
Due to the suppressed AMXs,  the soliton step is sufficiently long, allowing for direct access to the single soliton state using simple laser piezo frequency tuning \cite{Herr:14, Guo:16}, without any other complex tuning schemes or auxiliary lasers. 
An erbium-doped fiber amplifer (EDFA) is used to increase the on-chip pump power to approximately 55.7~mW to seed soliton formation.
Figure \ref{Fig:4}(a) shows the single soliton spectrum, which, as expected, does not show prominent dispersive wave features caused by AMXs. 
The single soliton spectrum fit shows a 3-dB bandwidth of 16.3~nm, corresponding to a pulse duration of 156~fs. 

We further characterize the single-sideband (SSB) phase noise of the K-band microwave carrier generated by photodetection of the soliton repetition rate, and investigate the phase noise reduction via operating the soliton at a ``quiet point'' \cite{Yi:17}. 
Figure \ref{Fig:4}(b) shows the soliton phase noise measurement with different pump laser detunings. 
The detuning values are measured using a vector network analyser (VNA) that probes the resonance frequency relative to the laser \cite{Guo:16}. 
The ``quiet point''~\cite{Yi:17} is a phenomenon that the soliton phase noise reaches its minimum at a particular detuning value. 
It is resulted from the balance of dispersive-wave-induced soliton spectral recoil and soliton-induced Raman self-frequency shift \cite{Karpov:16}.  
Different from our previous studies where operation at the quiet point enables more than 20 dBc/Hz phase noise reduction, here we only observe less than 4 dBc/Hz reduction measured within 1 kHz to 100 kHz Fourier offset frequency.  
To evidence the phase noise reduction at the quiet point, the soliton repetition rate shift and the phase noise value at 3.852 kHz Fourier offset frequency where the diode laser (Toptica CTL) exhibits a characteristic phase noise feature, are measured with different detuning values, as shown in Fig. \ref{Fig:4}(c).
It seems that the minimum phase noise is reached at 217.2 MHz detuning.  
The weak quite point effect agrees with the soliton spectrum that dispersive wave generation is inhibited, due to the suppressed spatial mode interaction and avoided-mode crossings. 

%%%%%%%%%%%%%%%%%%%%%%%%%
%%%%%%%%%%%%%%%%%%%%%%%%%

Previously, on integrated Si$_3$N$_4$ platform, single solitons of repetition rates below 20~GHz driven by CW pump have only been generated in microring resonators \cite{Liu:20}.
The microring resonator of 19.6~GHz FSR has a footprint of 4.2~mm$^2$, 20~times larger than our racetrack microresonator of 19.8~GHz FSR (0.21~mm$^2$).  
Such vast difference in device footprint is highlighted in Supplementary Information. 
The reduced footprint is critical for photonic integration of other functionalities, e.g. the monolithic integration of piezoelectric modulators \cite{Tian:20} on top of the racetrack microresonator, for soliton actuation \cite{Liu:20a} or laser frequency tuning via self-injection locking \cite{Lihachev:21}. 
Since the capacitance of piezoelectric modulators is proportional to the device area not perimeter, a small footprint allows to reduce the time constant and increase the modulation speed. 
In addition, the large aspect ratio of the racetrack microresonator,  length ($L=3500$ $\mu$m) to span ($d=60$ $\mu$m), facilitates the simulation and calculation of microresonator dispersion profile. 
This is because that the Euler bending sections only occupy a small portion of the entire racetrack, and the integrated microresonator properties are dominated by the straight waveguide sections. 
Finally, we emphasize that, such footprint reduction is enabled by using deep-ultraviolet (DUV) stepper lithography to pattern racetrack microresonators of high aspect ratio. % (60~$\mu$m span, 3500~$\mu$m length). 
In comparison, when using electron beam lithography (EBL), the writing field size limits the device shape and aspect ratio. 
Finger- or snail-shaped microresonators have been developed \cite{Xuan:16, Ye:21a} to confine the microresonator in one or few EBL writing fields, such that the impact of stitching errors is minimized \cite{Ye:21, Ji:19a}.  
For example, we note that, very recently, Ref.~\cite{Ye:21a} has shown 1~mm$^2$ device footprint for microresonators of 14.0~and 20.5~GHz FSR, and single soliton generation in these devices. 

%%%%%%%%%%%%%%%%%%%%%%%
%%%%% Conclusion %%%%%
%%%%%%%%%%%%%%%%%%%%%%%

In summary, we adapt \cite{Fujisawa:17, Vogelbacher:19}, optimize, and implement Euler bends to build compact racetrack microresonators based on ultralow-loss, multi-mode, Si$_3$N$_4$ photonic circuits. 
The optimized racetrack microresonator has a significantly reduced device footprint, critical for high device density and integration \cite{Roeloffzen:13}. 
It can be key building blocks for nonlinear photonic applications, such as microwave-repetition-rate soliton microcombs \cite{Liu:20, YuM:20}, travelling-wave optical parametric amplifiers \cite{Pu:18, Riemensberger:21, Ye:21a}, frequency conversion \cite{ChenJ:19}, or resonant electro-optic modulators \cite{Wang:18, Ahmed:19a} and frequency combs \cite{Zhang:19}.  
The adiabatic Euler bend is also useful for linear circuits based on beam splitters and interferometers that are widely used in integrated programmable processors \cite{Zhuang:15} and photonic quantum computing \cite{WangJW:20, Arrazola:21}. 
The simple design rules and algorithms illustrated here can be easily implemented in Si$_3$N$_4$ foundry process \cite{Munoz:19}. 

\begin{footnotesize}
\noindent \textbf{Funding Information}: 
This work was supported by the Air Force Office of Scientific Research (AFOSR) under Award No. FA9550-19-1-0250, 
by Contract HR0011-20-2-0046 (NOVEL) from the Defense Advanced Research Projects Agency (DARPA), Microsystems Technology Office (MTO), 
by the Swiss National Science Foundation under grant agreement No. 176563 (BRIDGE), and by the EU H2020 research and innovation programme under grant agreement No. 965124 (FEMTOCHIP). 

\noindent \textbf{Acknowledgments}: 
We thank Guanhao Huang and Miles Anderson for the fruitful discussion. 
The Si$_3$N$_4$ chips were fabricated in the EPFL center of MicroNanoTechnology (CMi). 

\noindent \textbf{Author contributions}: 
X.J.  and J.L.  performed the numerical simulation and analytical study. 
X.J.  and J.L.  designed the samples, with the assistance from Z.Q..
J.L.  and R.N.W. fabricated the samples.
X.J. and J.H. characterized the samples with assistance from J.R. and J.L..
J.H. performed the soliton generation and phase noise measurement with the assistance from J.R. and X.J.  
X.J. ,J.L. and J.H. analyzed the data.
J.L. and X.J.  wrote the manuscript with input from others.  
T.J.K. supervised the project.

%\noindent \textbf{Data Availability Statement}: The code and data used to produce the plots within this work will be released on the repository \texttt{Zenodo} upon publication of this preprint.

\end{footnotesize}
%\vspace{-0.3cm}
\bibliographystyle{apsrev4-1}
\bibliography{bibliography}
\end{document}

%\title{Supplementary Information for: Suppression of spatial mode interaction in ultralow-loss, multi-mode photonic integrated circuits}
\title{Supplementary Information for: Compact, spatial-mode-interaction-free, \\ultralow-loss, nonlinear photonic integrated circuits}
\author{Xinru Ji, Junqiu Liu, Jijun He \textit{et al.}}

\maketitle
%\pagebreak

%%%%%%%%%%%%%%%%%%%%%%%%%%%%%%%%%%%%%%%%%%%%%%%%%%%%%%%%%%%%%%%
\section{Mathematical derivation of partial Euler bends}
\vspace{0.5cm}

The partial Euler $\pi$-bend model mentioned in the main manuscript consists of two Euler splines and a circular bend, following Ref. \cite{Fujisawa:17}.  
In this section, the mathematical description of Euler splines and the geometric construction of partial Euler bends is presented in detail.

\begin{figure}[b!]
\centering
\renewcommand{\figurename}{Supplementary Figure}
\includegraphics[width=0.5\columnwidth]{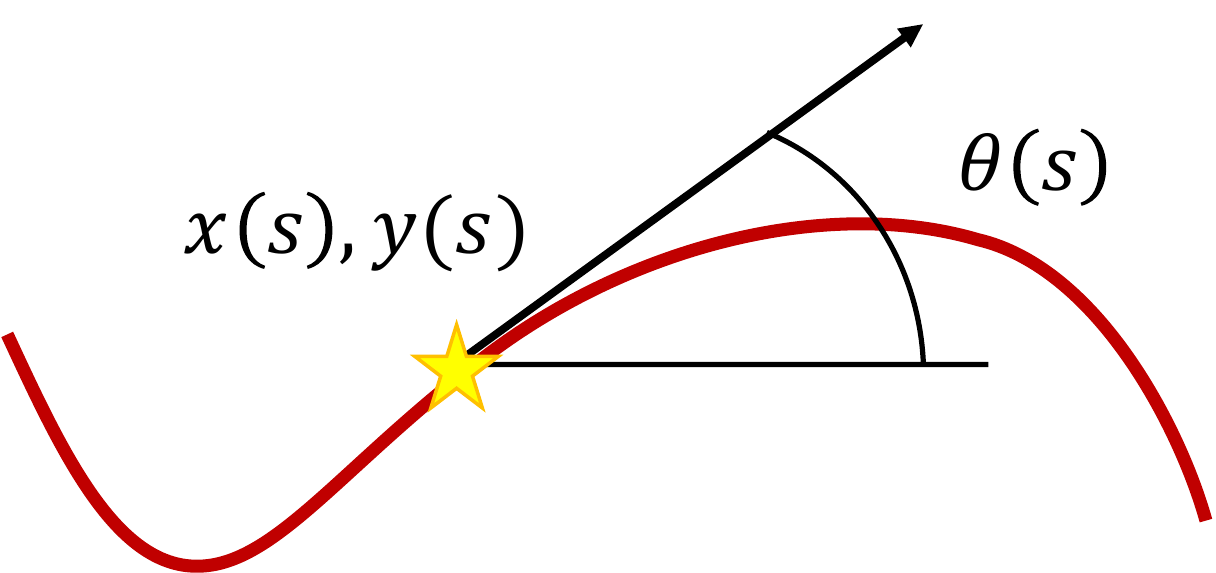}
\caption{Definition of a curve via tangent angle $\theta(s)$ along path parameter $s$}
\label{fig:1}
\end{figure}

\begin{figure}[b!]
\centering
\renewcommand{\figurename}{Supplementary Figure}
\includegraphics[width=0.5\columnwidth]{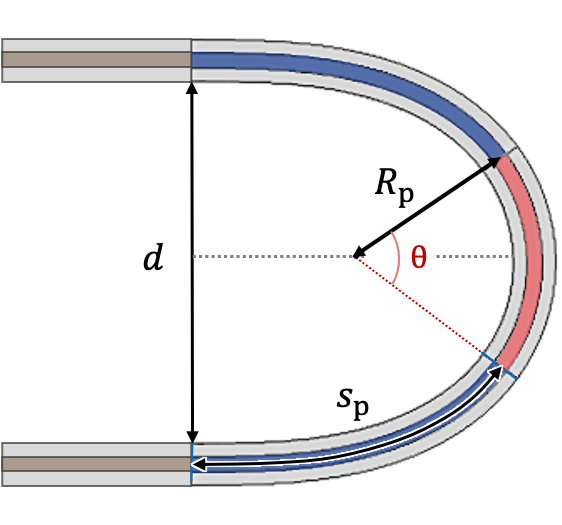}
\caption{Geometrical construction of partial Euler bend with portion $p=1-\frac{\theta}{\pi}$}
\label{fig:2}
\end{figure}

A curve can be defined by its tangent angle $\theta(s)$ and curvature $k(s)$ along its path length $s \ (s\geq0)$.  
In this definition, $k=\text{d}\theta(s)/\text{d}s$. 
The tangent angle $\theta(s)$ at the point $[x(s), y(s)]$ with path $s$ is illustrated in Supplementary Fig.~\ref{fig:1}, and can be calculated with integral $\theta(s)=\theta(0)+\int_{0}^{s} k(u)\text{d}u$.  
The coordinate in the Cartesian system $x$ -- $y$ is expressed as: 

\begin{equation}
\begin{aligned}
&x(s)=x(0)+\int_{0}^{s} \cos \theta(u)\text{d}u \\
&y(s)=y(0)+\int_{0}^{s} \sin \theta(u)\text{d}u
\end{aligned}
\end{equation}

An Euler spline is a curve whose bend curvature $k(s)$ increases linearly along its path length $s$, with the linear coefficient $\alpha$, i.e. $k(s)=\alpha \cdot s$. 
The coordinate $(x, y)$ can thus be expressed using Fresnel integrals:

\begin{equation}
\begin{aligned}
&x(s)=\int_{0}^{s} \cos \left(\frac{\alpha}{2} \cdot u^{2}\right) \text{d}u \\
&y(s)=\int_{0}^{s} \sin \left(\frac{\alpha}{2} \cdot u^{2}\right) \text{d}u
\end{aligned}
\label{eq:1}
\end{equation}

A partial Euler $\pi-$bend, as shown in Supplementary Fig.~\ref{fig:2}, is featured with an Euler portion factor $p$ that denotes the ratio of two Euler splines in the total $\pi$-bend. 
In this case, $p=0$ corresponds to a circular $\pi-$bend (with constant curvature), and $p=1$ corresponds full Euler $\pi-$bend. 
In the partial Euler $\pi-$bend, a circular bend is introduced with angle $\theta$.  
The relation of $\theta$ and $p$ is

\begin{equation}
p=1-\frac{\theta}{\pi}
\end{equation}

\begin{figure}[b!]
\centering
\renewcommand{\figurename}{Supplementary Figure}
\includegraphics[width=1\columnwidth]{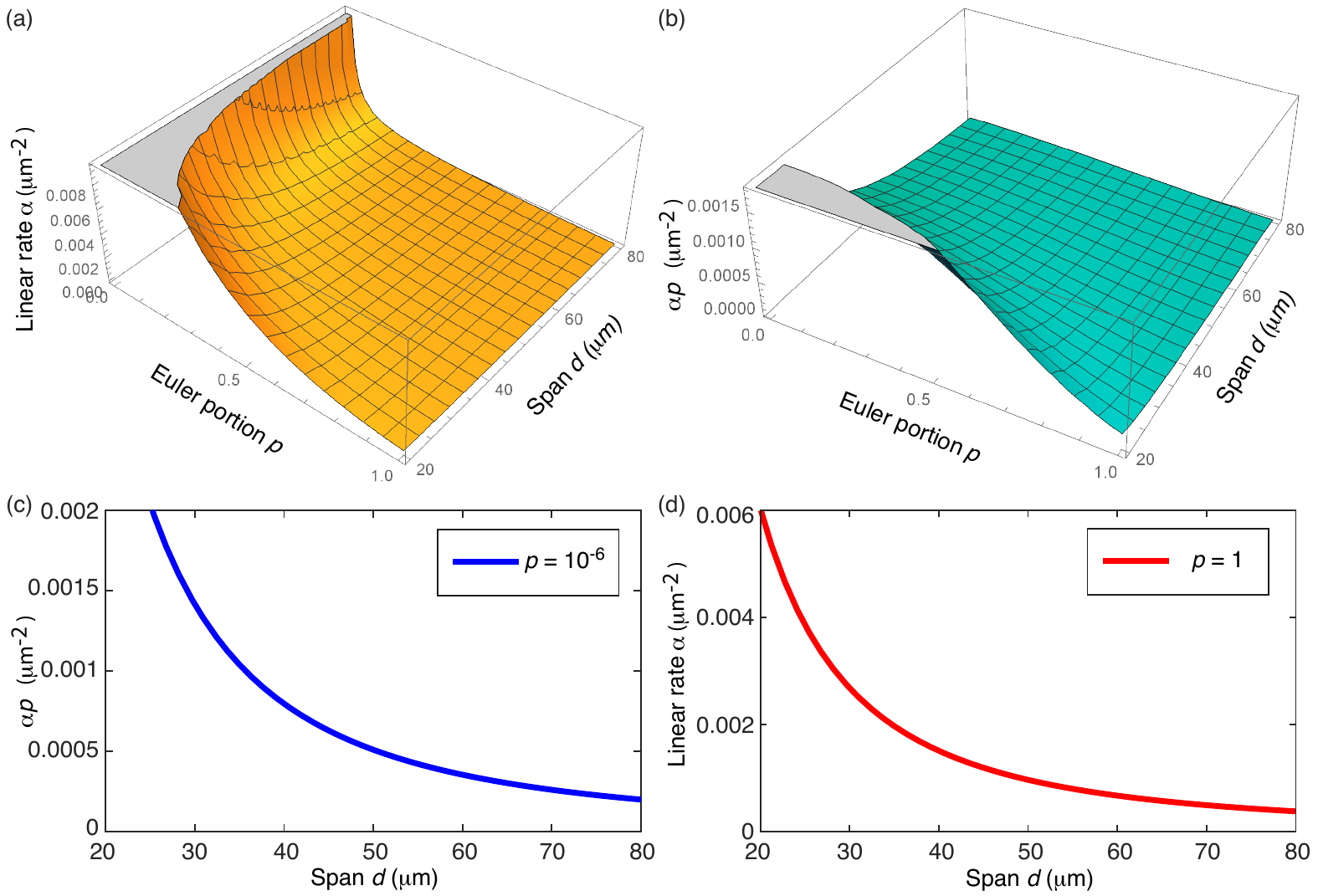}
\caption{Numerical study of the relation between $\alpha, p, d$. 
(a) 3D plot of $\alpha$ as a function of  $p\in(0,1]$ and $d\in[20, 80]$. 
(b) 3D plot of $\alpha p$ as a function of  $p\in(0,1]$ and $d\in[20, 80]$. 
(c) 2D plot of $\alpha p$ as a function of $d$, when $p\rightarrow0$ (here we use $p=10^{-6}$).  
In this case, the partial Euler bend degenerates to a full circular bend with a radius $R_p=d/2$ and $\alpha(p)p=4/\pi d^2$.
}
\label{fig:math}
\end{figure}

The total path length $s_p$ (marked in Supplementary Fig.~\ref{fig:2}) of one Euler bend from the straight waveguide to the circular bend of angle $\theta$ is derived as
\begin{equation}
\begin{aligned}
\frac{\text{d}\beta}{\text{d}s}&=\alpha s \\
\Rightarrow \beta&=\frac{\alpha s_p ^2}{2}\\
\Rightarrow s_p&=\sqrt{\frac{2\beta}{\alpha}}=\sqrt{\frac{\pi p}{\alpha}}
\end{aligned}
\end{equation}
where $\beta$ is the angle of the Euler bend, such that $\beta=(\pi-\theta)/2=\pi p/2$. 

The radius $R_p$ of the circular bend is derived as
\begin{equation}
\begin{aligned}
R_p=\frac{1}{\alpha s_p}=\frac{1}{\sqrt{\alpha\pi p}}
\end{aligned}
\end{equation}

%The path length of one of the two Euler bends, $s_p=\sqrt{p \theta/\alpha}$, can be calculated from Fresnel integral Eqs. \ref{eq:1}. 

The linear increase rate $\alpha$ in Eqs. \ref{eq:1} determines the footprint (i.e. the geometric area) of a partial Euler $\pi-$bend. 
However, when the values of $p$ and the span of the $\pi$-bend $d$ are given, $\alpha$ can be uniquely determined. 
Using Eqs. \ref{eq:1} and the geometry shown in Supplementary Fig. \ref{fig:2}, the relation of $\alpha, p, d$ is 
\begin{equation}
\begin{aligned}
\frac{d}{2}&=y(\sqrt{\frac{\pi p}{\alpha}})+\frac{\sin\frac{\pi(1-p)}{2}}{\sqrt{\alpha \pi p}}\\
\Rightarrow \alpha&=\frac{[2\sqrt{2}\int_{0}^{\sqrt{\pi p/2}} \sin t^2 \text{d}t+\frac{2}{\sqrt{\pi p}}\sin \frac{\pi(1-p)}{2}]^2}{d^2}
\end{aligned}
\end{equation}

Supplementary Figure \ref{fig:math}(a, b) presents the 3D plot of $\alpha$ and $\alpha p$ as a function of  $p\in(0,1]$ and $d\in[20, 80]$. 
Supplementary Figure \ref{fig:math}(c) presents the 2D plot of $\alpha p$ as a function of $d$, when $p\rightarrow0$ (here we use $p=10^{-6}$, as $p=0$ causes numerical divergence).  
When $p\rightarrow0$, $\theta\rightarrow\pi$, the partial Euler bend degenerates to a full circular bend with a radius $R_p=d/2$, and
\begin{equation}
\begin{aligned}
\frac{1}{\sqrt{\alpha\pi p}}&=\frac{d}{2}\\
\Rightarrow \alpha p&=\frac{4}{\pi d^2}
\end{aligned}
\end{equation}
In this case, the Euler bend length is infinitely small, i.e. $s_p(p\rightarrow0)\rightarrow0$. 
However, to change the radius from infinitely large ($+\infty$) to a finite value of $d/2$ with $s_p\rightarrow0$, the linear changing rate is infinity, i.e. $\alpha(p\rightarrow0)\rightarrow+\infty$, but yielding a finite value of $\alpha(p)p=4/\pi d^2$.
Supplementary Figure \ref{fig:math}(d) presents the 2D plot of linear rate $\alpha$ as a function of $d$ in the case with a full Euler bend $p=1$.  
As expected, larger $d$ requires smaller $\alpha$ to form a $\pi$-bend to connect the two straight waveguide. 

\begin{figure}[b!]
\centering
\renewcommand{\figurename}{Supplementary Figure}
\includegraphics[width=1\columnwidth]{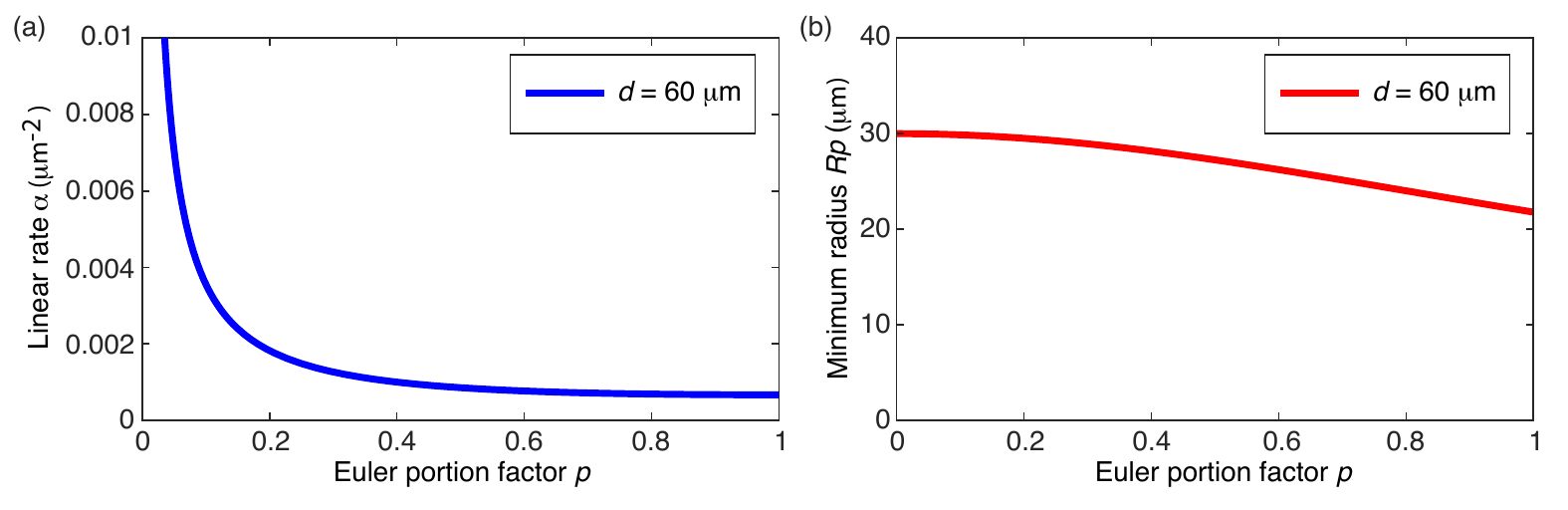}
\caption{Numerical study of the relation between $\alpha, R_p, p$. 
(a) 2D plot of $\alpha$ as a function of  $p\in(0,1]$ with a constant $d=60$ $\mu$m. 
(b) 2D plot of $R_p$ as a function of  $p\in(0,1]$ with a constant $d=60$ $\mu$m. 
}
\label{fig:math2}
\end{figure}

Next, we fix $d=60$ $\mu$m that is used in our experiment, and numerically study the relation of linear rate $\alpha$ and minimum bending radius $R_p=(\alpha\pi p)^{-1/2}$ (that is also the radius of the circular bend) as a function of the Euler portion factor $p$, as shown in Supplementary Fig. \ref{fig:math2}. 
The numerical plot of $\alpha$ as a function of $p$ with a constant $d$ in Supplementary Fig.~\ref{fig:math2}(a) shows that, the value of $\alpha$ decreases monotonously with $p$, and its minimum value is reached when $p=1$. 
Also, $R_p$ decreased monotonously with $p$, as shown in Supplementary Fig.~\ref{fig:math2}(b). 
When $p=0$, the partial Euler bend degenerates to the circular bend where $R_p=d/2=30$ $\mu$m.  
%From these two plots, it becomes clear that the portion of circular bend in the partial Euler bend is to prevent large $\alpha$ and small $R_p$. 

It seems that the linear rate $\alpha$ of bending curvature change is a parameter representing adiabaticity, and the lowest value of $\alpha$ is reached when $p=1$.  
However, questions remain and further analysis is needed. 
For example, Ref. \cite{Chen:12} illustrates that the essential parameter determining the adiabaticity is the changing rate of mode effective refractive index that can be numerically analyzed via conformal mapping. 

\clearpage

%%%%%%%%%%%%%%%%%%%%%%%%%%%%%%%%%%%%%%%%%%%%%%%%%%%%%%%%%%%%%%%
%\vspace{0.5cm}
\section{Microresonator transmission spectra}
\vspace{0.5cm}

The normalised transmission traces for $d=60$ $\mu$m span racetrack microresonators with Euler bends ($p=1$), partial Euler bends ($p=0.6$), and circular bends ($p=0$),  are shown in Supplementary Fig.~\ref{fig:trans}. 
Transmission traces from two lasers, spanning from 184 to 200 THz (1500 to 1630 nm) and from 199 to 221 THz (1350 to 1505 nm), are stitched together to achieve broadband measurement. 

Experimentally, the improvement in device performance is prominent. 
In the racetrack microresonator with circular bends ($p=0$),  avoided mode crossings (AMX) are evident, yielding in a modulation in the transmission background and a significant loss rate for some resonances. 
In comparison, the racetrack microresonator with partial Euler bends ($p=0.6$) is less impacted by AMX. 
Nonetheless, such modulation disappears in the racetrack microresonators with Euler bends ($p=1$), as the continuous transmission envelope encounters almost no abrupt change. 
The wavelength-dependent resonance extinction ratio is due the the microresonator's external coupling strength with the bus waveguide which is wavelength-dependent \cite{Cai:00}.

\begin{figure}[b!]
\centering
\renewcommand{\figurename}{Supplementary Figure}
\includegraphics[width=0.8\columnwidth]{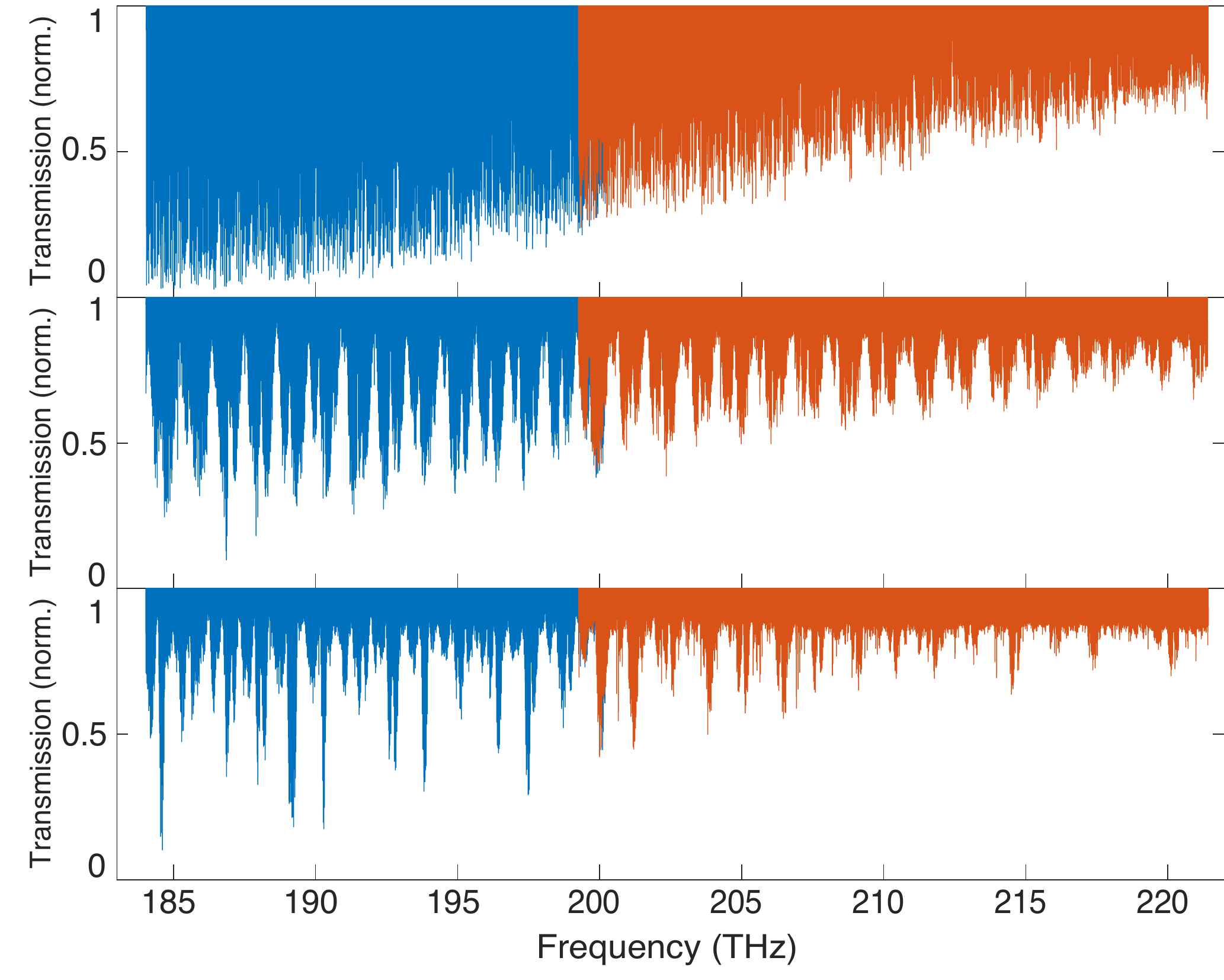}
\caption{Optical transmission spectra of 19.8-GHz-FSR racetrack microresonators with Euler bends ($p=1$, top), partial Euler bends ($p=0.6$, middle), and circular bends ($p=0$, bottom). 
All racetrack microresonators have the same span of $d=60 \mu$m for the $\pi$-bends.}
\label{fig:trans}
\end{figure}

The inter-modal modulation pattern in the transmission spectra of racetrack microresonators with circular and partial Euler bends arises from the interference between TE$_{00}$ mode and higher-order modes (TE$_{10}$, TE$_{20}$, etc). 
The wavelength-dependent period of transmission window is linked to the group index difference between TE$_{00}$ mode and other higher-order TE modes.
For example, the destructive interference condition between TE$_{00}$ mode and TE$_{10}$ mode writes:

\begin{equation}
\Delta \phi=\phi_{10}-\phi_{00}=\frac{\Delta S}{c} \omega\left(n_{10}-n_{00}\right)
\end{equation}

The frequency interference period is then:

\begin{equation}
\begin{aligned}
\frac{\Delta \phi}{\Delta \omega} & \simeq\frac{\Delta S}{c}\left(n_{10}+\frac{\partial n_{10}}{\partial \omega}\cdot \omega-n_{00}-\frac{\partial n_{00}}{\partial \omega}\cdot \omega\right)\\
&=\frac{\Delta S}{c}\left(n_{g, 10}-n_{g, 00}\right) \\
\Rightarrow \Delta f&=\frac{c}{\Delta S \cdot \left(n_{g, 10}-n_{g, 00}\right)}
\end{aligned}
\label{eq:2}
\end{equation}
where $n_{00}$ ($n_{g, 00}$) and $n_{10}$ ($n_{g, 10}$) are the effective (group) indices of the TE$_{00}$  and TE$_{10}$ modes, $(n_{g, 10}-n_{g, 00})\Delta S$ is the optical path difference between the TE$_{00}$ and TE$_{10}$ mode. $\Delta S$ is set to 7000 $\mu$m, corresponding to two times of the straight waveguide length ($L=3500$ $\mu$m), as the bending section is negligible compared to the straight section ($L_{\text{bend}}/L_{\text{straight}} \sim 5 \%$).

The FEM simulation results of group indices for TE$_{00}$, TE$_{10}$ and TE$_{20}$ modes, as well as the wavelength-dependent destructive interference period $\Delta f$, are shown in Supplementary Fig. \ref{fig:Multimode}. 
This transmission analysis is only applied to the racetrack microresonators with partial Euler bend, as the increased complexity with the circular bend makes the analysis challenging.
The double-hoof shape of the interference fringes in the transmission spectrum are due to the presence of both TE$_{10}$ and TE$_{20}$ modes. 
From the simulation result shown in Supplementary Fig. \ref{fig:Multimode}(b), at 190 THz, the destructive interference periods for TE$_{00}$-TE$_{10}$ modes and TE$_{00}$-TE$_{20}$ modes are $\sim$0.58 THz and $\sim$0.24 THz, respectively. 
In addition, both periods increase with optical frequency.
These results qualitatively agree with the experimentally measured microresonator transmission spectrum shown in Supplementary Fig. \ref{fig:trans} middle, where $\Delta f_{00,10} \sim$ 1 THz and $\Delta f_{00,20} \sim$ 0.4 THz. 
The mismatch is likely caused by the bending section, whose path length and mode group indices should also be included in Eq. \ref{eq:2}.

\begin{figure}[b!]
\centering
\renewcommand{\figurename}{Supplementary Figure}
\includegraphics[width=0.8\columnwidth]{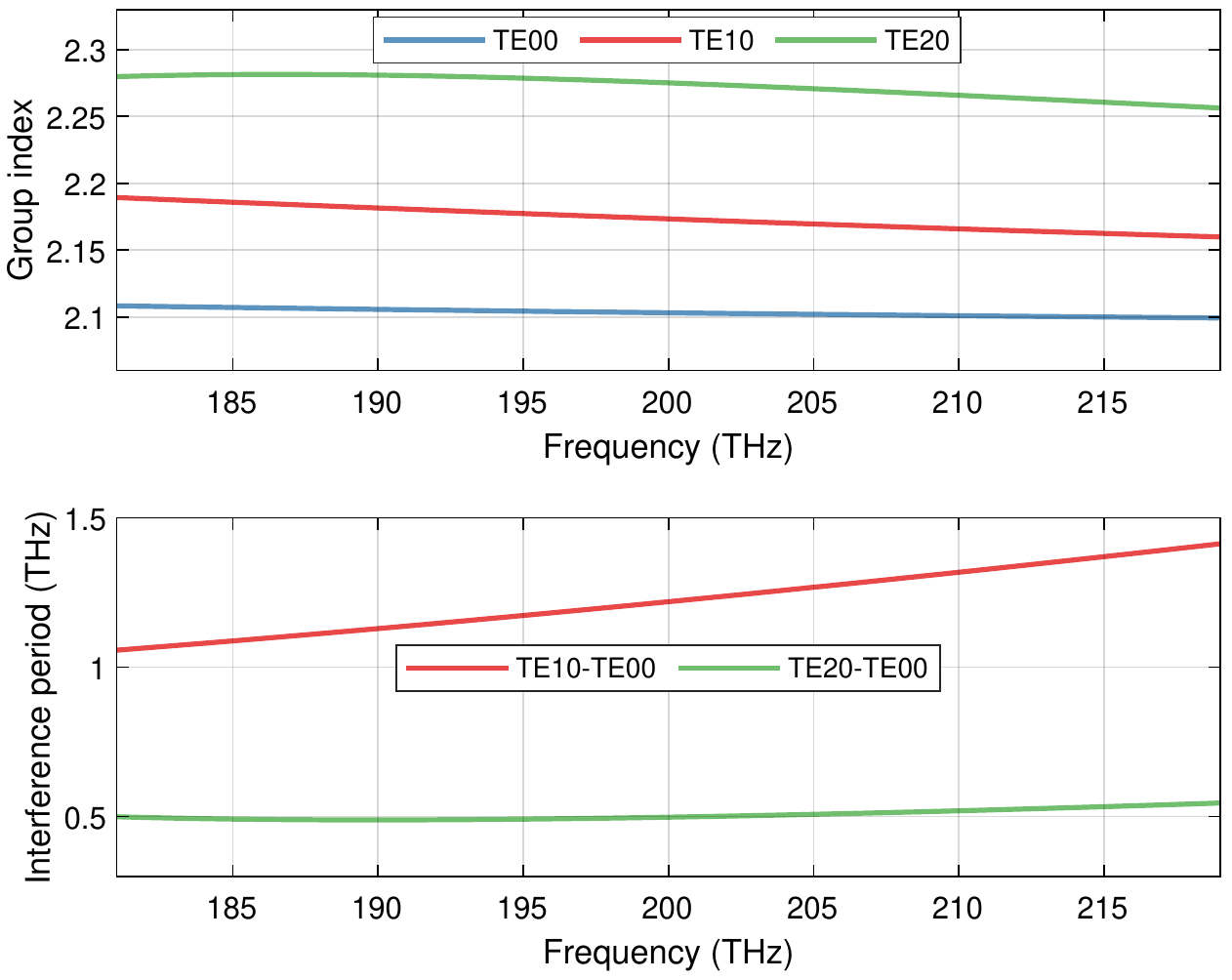}
\caption{FEM simulation for inter-modal modulation period analysis. 
(a) FEM simulation of group indices $n_{g, 00}$, $n_{g, 10}$ and $n_{g, 20}$ for TE$_{00}$, TE$_{10}$ and TE$_{20}$ eigenmode of $2.2\times0.95$ $\mu$m$^2$ straight Si$_3$N$_4$ waveguide.(b) Multi-mode interference period between TE$_{00}$-TE$_{10}$ and TE$_{00}$-TE$_{20}$ mode, calculated with Eq.\ref{eq:2}.}
\label{fig:Multimode}
\end{figure}

\clearpage

%%%%%%%%%%%%%%%%%%%%%%%%%%%%%%%%%%%%%%%%%%%%%%%%%%%%%%%%%%%%%%%
%\vspace{0.5cm}
\section{Racetrack design layout}
\vspace{0.5cm}

\begin{figure*}[b!]
\renewcommand{\figurename}{Supplementary Figure}
\centering
\includegraphics{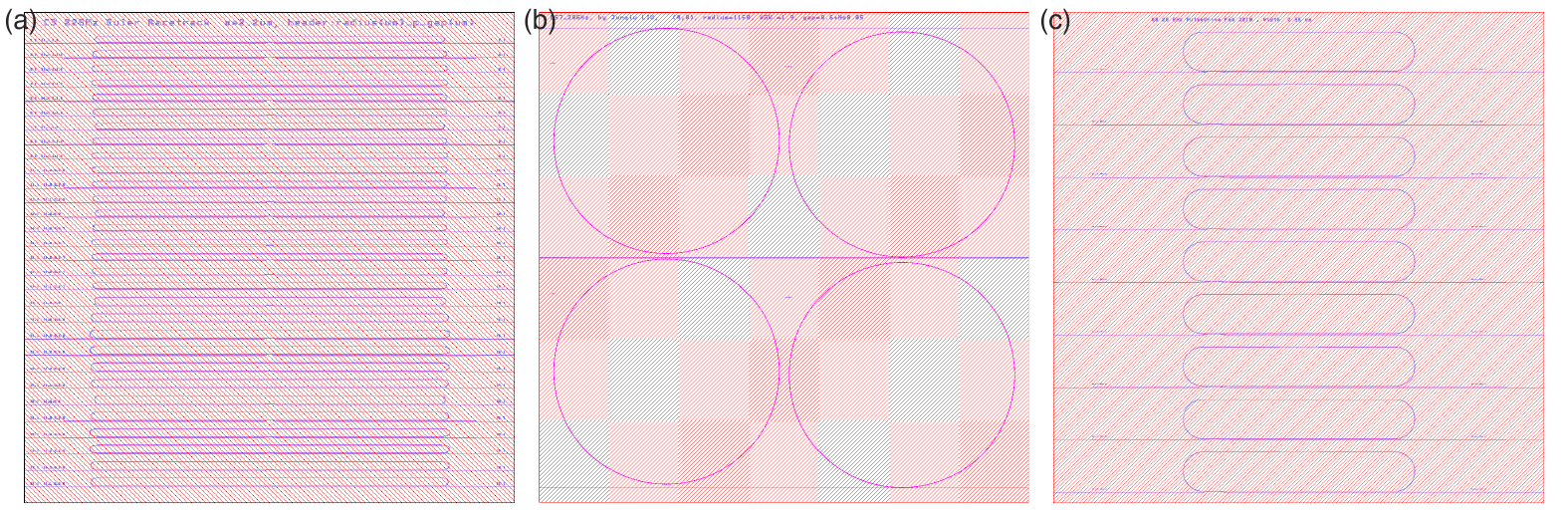}
\caption{GDS design layouts of racetrack microresonators with Euler bends and around 19.8 GHz FSR (a), microring resonators (b), and racetrack microresonators with circular bends and 28 GHz FSR (c), on $5\times5$ mm$^2$ chips. 
The racetrack microresonators with Euler bends, partial Euler bends, and circular bends, which are studied in the main manuscript, are the 7th, 10th, and 12th devices counted from the chip top in (a). }
\label{fig:gds}
\end{figure*}

Supplementary Figure \ref{fig:gds} shows the GDS design layouts of different Si$_3$N$_4$ microresonators of FSR varying from 19 to 28 GHz FSRs, on $5\times5$ mm$^2$ chips. 
The microresonator is coupled to a bus waveguide whose waveguide width is identical to the microresonator's waveguide width, to achieve high coupling ideality\cite{Pfeiffer:17b}. 
Supplementary Figure \ref{fig:gds}(a) shows the chip containing the racetrack microresonators with Euler bends, partial Euler bends, and circular bends. 
As is seen here, there are 30 racetrack microresonators of around 19.8 GHz FSR. 
The racetrack microresonators with Euler bends, partial Euler bends, and circular bends, which are studied in the main manuscript, are the 7th, 10th, and 12th devices counted from the chip top. 
Supplementary Figure \ref{fig:gds}(b) shows the chip containing four microring resonators of 19.6 GHz FSR, which has been used in Ref. \cite{Liu:20}. 
Supplementary Figure \ref{fig:gds}(c) shows the chip containing nine racetrack microresonators with circular bends and 28 GHz FSR,  which has been used in Ref. \cite{Anderson:21}. 
Despite the FSRs are similar, the difference in device footprint is vast. 
Apparently, our racetrack microresonators with Euler bends have significantly reduced footprint compared to our previous studies. 

%%%%%%%%%%%%%%%%%%%%%%%%%%%%%%%%%%%%%%%%%%%%%%%%%%%%%%%%%%%%%%%
\vspace{0.5cm}
\section{Finite-element simulations}
\vspace{0.5cm}

In addition to FDTD simulations shown in the main text, here we also use finite-element method (FEM) with \textit{COMSOL Multiphysics} to simulate 3D light propagation profile at optical wavelength of 1550 nm in the partial Euler bend. 
We use FEM to find the optimum value of Euler portion factor $p$ that shows best adiabatic performance, namely highest transmission of the TE$_{00}$ mode over the $\pi$-bend. 

In the simulation, we employ the inherent symmetry of the transverse-electric (TE) modes to reduce the computation of field propagation to the half of the model volume, with a perfect magnetic conductor boundary condition. 
This significantly reduces the memory requirements and computation time.  
Moreover, the simulation is performed with direct MUMPS solver in the out-of-core mode, enabling partial offload of the problem to the hard disk for less memory consumption.

\begin{figure}[t!]
\centering
\renewcommand{\figurename}{Supplementary Figure}
\includegraphics[width=1\columnwidth]{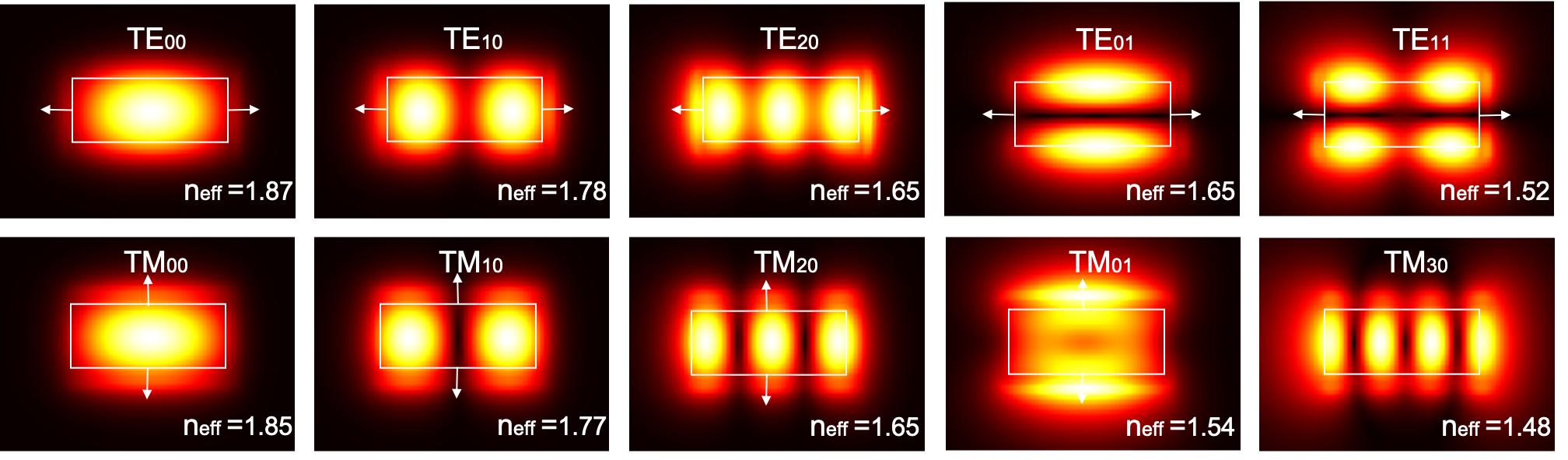}
\caption{TE and TM eigenmodes of the Si$_3$N$_4$ waveguide with $2.2\times0.95$ $\mu$m$^2$ cross-section.}
\label{fig:eigenmode}
\end{figure}

\begin{figure}[b!]
\centering
\renewcommand{\figurename}{Supplementary Figure}
\includegraphics[width=1\columnwidth]{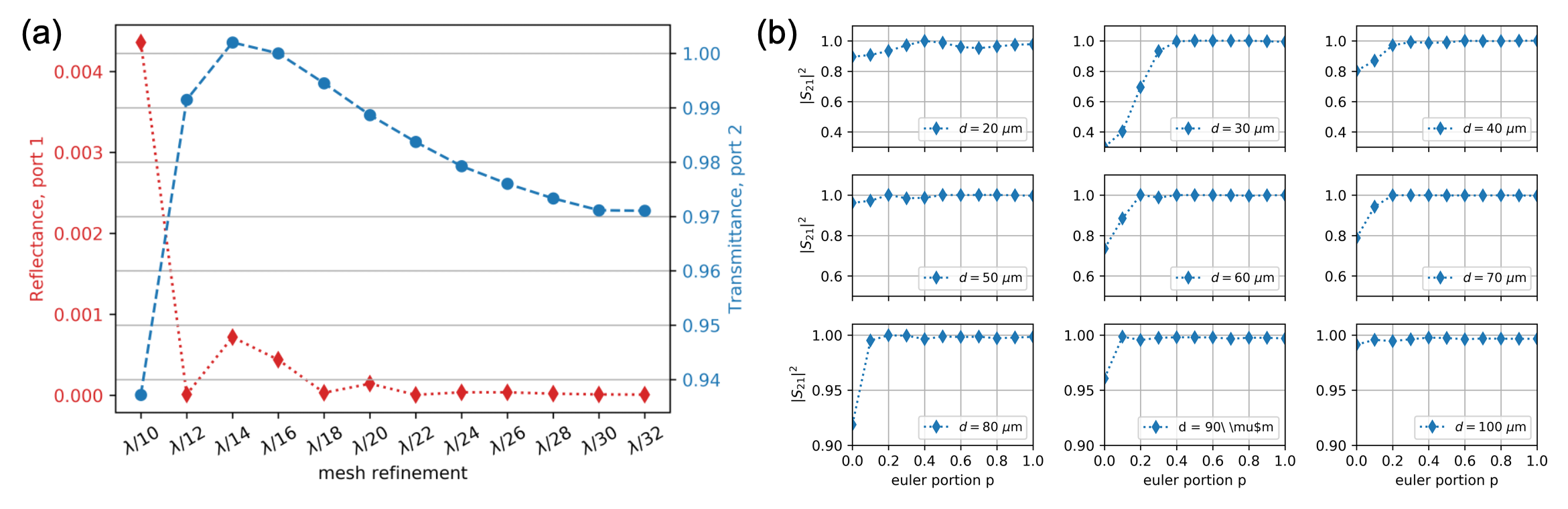}
\caption{(a) Convergence test result at different mesh refinements, in a model with parameters $d=60$ $\mu$m, $p=0.2$.  
Both the reflectance at port 1 and transmittance converge at port 2 converge for a mesh finer than $\lambda/30$, this value is then considered to yield valid results in simulations. 
(b)Transmission results for $\pi$-bends with Euler portion $p \in[0,1]$ and span $d \in[10,50]$ $\mu$m.
}
\label{fig:simulation}
\end{figure}

The studied Euler bend Si$_3$N$_4$ waveguide core is 2.2 $\mu$m wide and 0.95 $\mu$m high, and is fully cladded with SiO$_2$. 
A straight Si$_3$N$_4$ waveguide of this cross-section supports 10 transverse eigenmodes, as shown in Supplementary Fig.~\ref{fig:eigenmode}.
In the simulation, the TE$_{00}$ mode is launched at 1550 nm wavelength at the input port of the partial Euler bend with normalised power of unity.  
The transmittance at the output port on the other side of the $\pi$-bend is recorded, which is used to assess the adiabaticity of mode conversion.

The modeling of full-wave propagation is based on direct discretization of Maxwell’s equations. 
The physics interface employed here is the \textit{Wave Optics Module,} in the \textit{Electromagnetic Waves Frequency Domain}. 
This interface solves the time-harmonic wave equation for the electric field, and enables efficient simulation.  

The mesh size of the model is important for accuracy.  
As the Nyquist frequency indicates that the time-step size should be smaller than half the period of the highest frequency wave,  the size of one mesh element must not exceed half the wavelength of the propagating wave.  
Typically, it requires between five and ten data points per wavelength to obtain reasonable accuracy. 
In order to save computational time and ease memory requirement, meshes for the core and the cladding are built with different levels of refinement. 
For the waveguide core, the element size is set to be $\lambda$/30 ($\lambda=$1550 nm), which corresponds to 12 data nodes per wavelength and gives reliable results, as is confirmed by a convergence test in Supplementary Fig.~\ref{fig:simulation}(a). 
The remaining of the waveguide is filled with coarser meshes that originate at the core waveguide’s border and expand exponentially at a rate of 1.1. 
The complete mesh is built by sweeping a meshed surface from source to destination face.

Subsequently, three studies were conducted for the selected interface: 
two \textit{Boundary Mode Analysis} to compute the propagation constants or wave numbers, as well as the mode propagation profile, for a given frequency at the two ports, respectively; 
a following \textit{Frequency Domain} study step estimates the transmissive and reflective responses to the harmonic excitation, accounting for the effects of all eigenmodes that are properly resolved by the mesh in the full geometry. 
The resulting fields are normalized with respect to the integral of the power flow across each port cross-section.
In the \textit{Results} panel, the output port transmission is recorded by the scattering parameters ($S$-parameter) in terms of the power flow:

\begin{equation}
\begin{aligned} 
S_{11} &=\sqrt{\frac{\text { Power reflected from port } 1}{\text { Power incident on port l }}} \\ 
S_{21} &=\sqrt{\frac{\text { Power delivered to port } 2}{\text { Power incident on port } 1}} 
\end{aligned}
\end{equation}
where the amount of power flowing out of a port is given by the normal component of the Poynting vector $\mathbf{n} \cdot \frac{1}{2} \operatorname{Re}\left(\mathbf{E} \times \mathbf{H}^{*}\right)$. 
The transmittance and reflectance are interpreted as abs($S_{11}$)$^2$ and abs($S_{21}$)$^2$. 
In particular, finding the absolute value of the $S$-parameter is essential. 
Since both $S_{11}$ and $S_{21}$ are displayed in their complex representation, only the absolute value signifies the physical parameter of interest.

\begin{figure}[b!]
\centering
\renewcommand{\figurename}{Supplementary Figure}
\includegraphics[width=1\columnwidth]{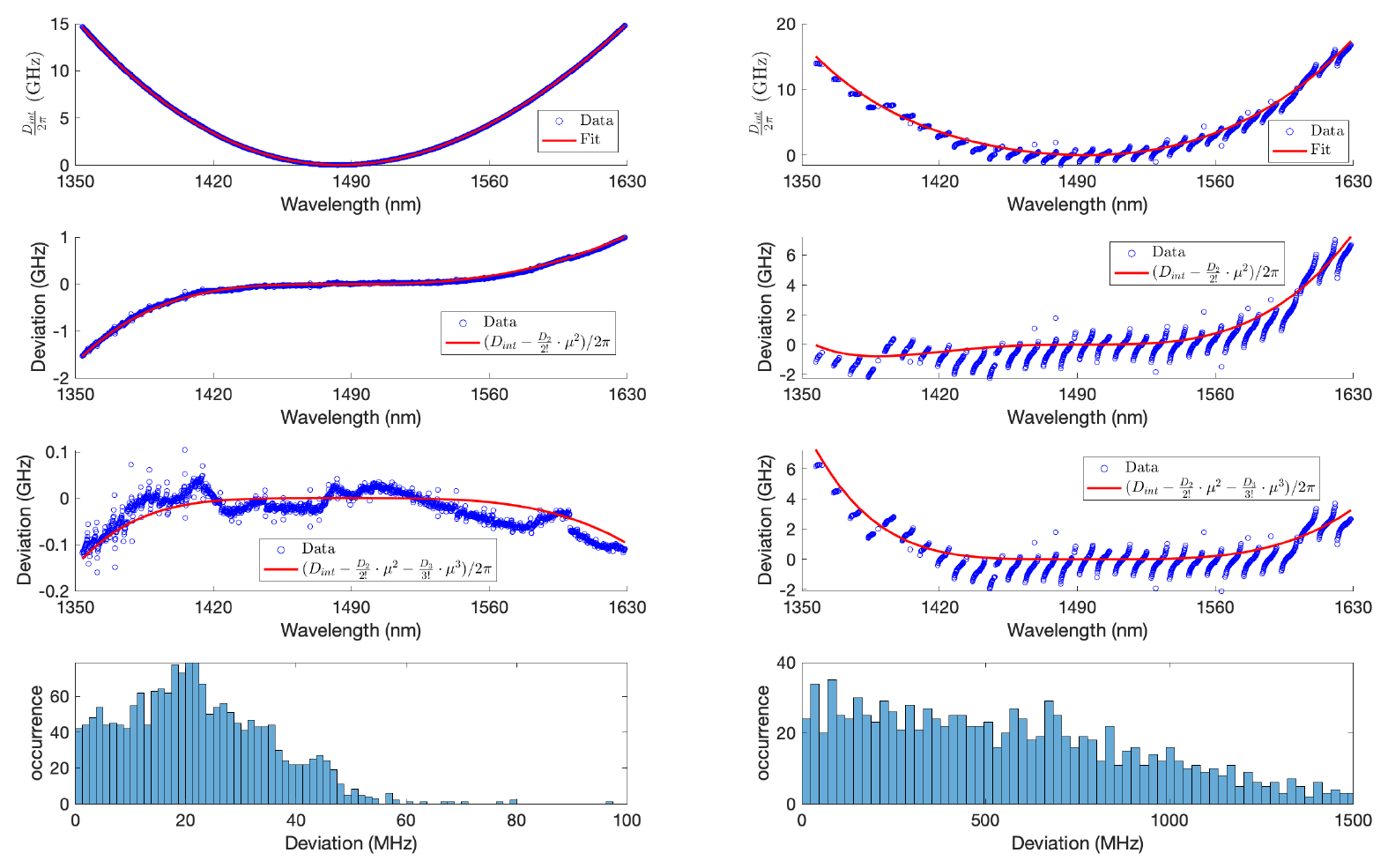}
\caption{Microresonator dispersion characterization of racetrack microresonators with Euler bends ($p=1$, left), and partial Euler bends ($p=0.6$, right). 
The measured integrated microresonator dispersion fitted with $D_\text{int}(\mu)=D_2\mu^2/2+D_3\mu^3/6+D_4\mu^4/24$ (top row), the resonance frequency deviations from $D_3\mu^3/6+D_4\mu^4/24$ (2nd row) and $D_4\mu^4/24$ (3rd row), and the histogram of frequency deviations from the $D_4\cdot\mu^4/24$ curve (bottom row). 
}
\label{fig:F5}
\end{figure}

\begin{figure}[b!]
\centering
\renewcommand{\figurename}{Supplementary Figure}
\includegraphics[width=1\columnwidth]{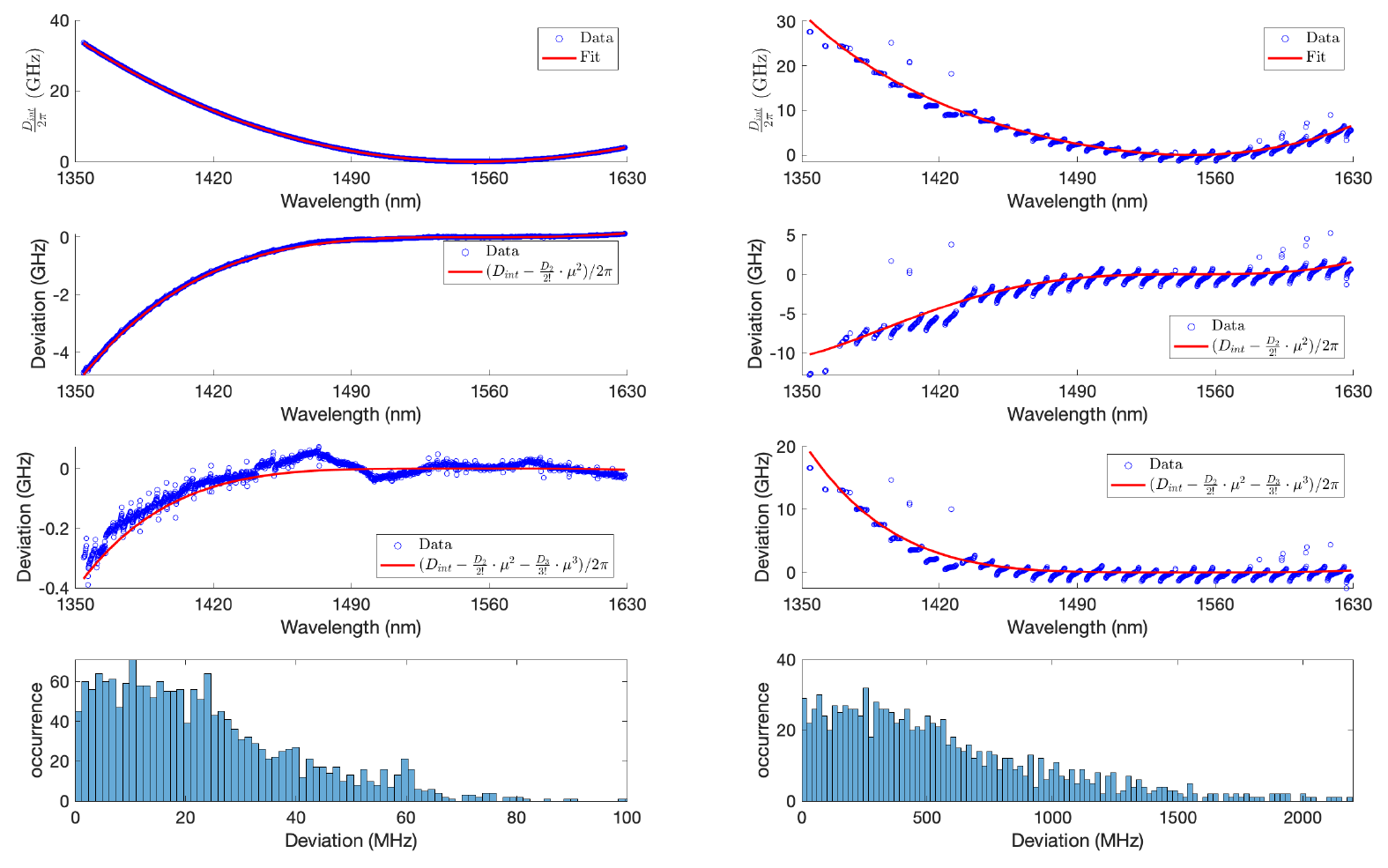}
\caption{Microresonator dispersion characterization of racetrack microresonators with Euler bends ($p=1$, left), and partial Euler bends ($p=0.6$, right). 
The measured integrated microresonator dispersion fitted with $D_\text{int}(\mu)=D_2\mu^2/2+D_3\mu^3/6+D_4\mu^4/24$ (top row), the resonance frequency deviations from $D_3\mu^3/6+D_4\mu^4/24$ (2nd row) and $D_4\mu^4/24$ (3rd row), and the histogram of frequency deviations from the $D_4\cdot\mu^4/24$ curve (bottom row). 
}
\label{fig:F6}
\end{figure}

The transmission results for the $\pi$-bend with dimension varying from $d=20$ to $100$ $\mu$m, are presented in Supplementary Fig.~\ref{fig:simulation}(b). 
Notably, the partial Euler bend (including the full Euler bend) has lower power loss and higher single mode conservation than the circular bend, particularly for the small $d$.  
Taking into consideration the numerical error, the difference in $\left|S_{21}\right|^{2}$ between a circular bend and a partial Euler bend diminishes for larger $d$, which is consistent with the fact that radiation loss decreases with increasing device size. 
Furthermore, contrary to the findings in Ref. \cite{Vogelbacher:19}, the simulation results indicate no evident optimal Euler portion for $\pi$-bends.  
This is presumably resulted from a wider waveguide core of 2.2 $\mu$m, which already confines the majority of the mode, leading to indiscernible loss between different geometries. 
Additionally, if the total loss is more or less the same for different Euler coefficients $p$, the geometry change of the produced device will not be a deciding factor in device performance, easing the fabrication requirement. 
The fluctuations in $\left|S_{21}\right|^{2}$ are seemingly stronger for small $d$, as $p$ changes from 0 to 1 , implying that an even finer mesh may be required to produce a valid result. 

Simulation performed here shows bending losses at 1550 nm wavelength for circular bends, partial Euler bends and Euler bends of various dimensions. 
In principle, it only provides limited information and does not predicts the broadband transmission response in Supplementary Figure \ref{fig:trans}, as higher-order eigenmodes may not be excited at this wavelength.
However, the FEM simulation allows simple pre-evaluations of device performance and serves as a reference in design.

%%%%%%%%%%%%%%%%%%%%%%%%%%%%%%%%%%%%%%%%%%%%%%%%%%%%%%%%%%%%%%%%
\vspace{0.5cm}
\section{Extra dispersion measurement data on other samples}
\vspace{0.5cm}

In addition to the three different racetrack microresonators shown in Fig. 2 and 3 in the main manuscript, here we provide extra measurement data on other samples. 
Since we use DUV stepper lithography that uniformly exposes the reticle pattern (containing $4\times4$ chips) over the full 4-inch wafer scale in discrete fields, we have multiple replica of the same designs at different wafer position. 
Our DUV stepper lithography's exposure array is found in Ref. \cite{Liu:21}. 

Supplementary Figures \ref{fig:F5} and \ref{fig:F6} show another two pairs of racetrack microresonators with $p=1$ and $p=0.6$ (four devices in total).
In each figure, the two racetrack microresonators are on the same chip, as mentioned in the previous section. 
Here we focus on the dispersion analysis of these new samples, as their statistical $Q$ values are similar to those of the samples shown in the main manuscript. 
In Supplementary Fig.~\ref{fig:F5}, the 90\% confidence interval of resonance frequency deviations from $D_4\mu^4/24$ curve is 40 MHz for the racetrack microresonator of Euler bends ($p=1$), and 1100 MHz for partial Euler bends ($p=0.6$). 
In Supplementary Fig.~\ref{fig:F6}, the 90\% confidence interval is 49.2 MHz for Euler bends ($p=1$), and 1313 MHz for partial Euler bends ($p=0.6$). 
The observations are same as that have been shown in the main manuscript.  

%%%%%%%%%%%%%%%%%%%%%%%%%%%%%%%%%%%%%%%%%%%%%%%%%%%%%%%%%%%%%%

%%%%%%%%%%%%%%%%%%%%%%%%%%%%%%%%%%%%%
%\vspace{0.5cm}
\clearpage
\section*{Supplementary References}
%\bigskip
\bibliographystyle{naturemag}
\bibliography{bibliography}